\documentclass[aps,prb,reprint,superscriptaddress]{revtex4-2}
\usepackage{physics}
\usepackage{bm,amsmath,amssymb,amsfonts,ulem}
\usepackage{graphicx}
\usepackage{xcolor}

\usepackage{hyperref}
\hypersetup{
	citecolor = blue,
	colorlinks = true,
	urlcolor = blue
}

\allowdisplaybreaks[4]

\begin{document}


\title{Theory of Nonlinear Hall Effect Induced by Electric Field and Temperature Gradient in 
	3D Chiral Magnetic Textures}
\author{Terufumi Yamaguchi} 
\author{Kazuki Nakazawa} 
\affiliation{RIKEN Center for Emergent Matter Science, 2-1 Hirosawa, Wako, Saitama 351-0198, Japan}
\author{Ai Yamakage} 
\affiliation{Department of Physics, Nagoya University, Nagoya 464-8602, Japan}
\date{\today}



\begin{abstract}
    We investigate the current proportional to the cross product 
	of an electric field and a temperature gradient,
	which we call the Nonlinear Chiral Thermo-Electric (NCTE) Hall effect,
	in 3D chiral magnetic structures.
	We analyze both discrete and continuous magnetization cases.
	From the analysis of discrete magnetization, 
	we clarify the conditions for the emergence of the NCTE Hall effect caused by magnetic textures.
	In the continuous magnetization case, 
	we show that the NCTE Hall conductivity is proportional to the density of magnetic monopoles.
	This study discovers unconventional phenomena unique to 3D magnetic textures, 
	and has significant implications for spintronics and quantum transport.
\end{abstract}
\maketitle



\textit{Introduction.--}
The exploration of nonlinear transport phenomena~\cite{Sodemann2015-qw,Mikhailov2016-rn,Hidaka2018-vb,Ma2019-wh,Nakai2019-sk,
Nandy2019-hw,Zeng2019-sr,Ishizuka2020-uu,Toshio2020-ax,Michishita2021-rn,Du2021-nc,Du2021-tv,Okumura2021-vz,
Michishita2022-ql,Du2021-nc,Liu2022-ei,Yamaguchi2024-yo,Arisawa2024-ot}
became a central focus of recent research in condensed matter physics, 
leading to the discovery of new phenomena that cannot occur in linear responses.
One such phenomenon is the nonlinear Hall effect~\cite{Sodemann2015-qw,Ma2019-wh,Nandy2019-hw,Du2021-nc,Du2021-tv}, 
where a current flows in a direction perpendicular to an applied electric field, 
similar to the conventional (anomalous) Hall effect~\cite{Karplus1954-bx,Sinitsyn2008-qh,Nagaosa2010-nq}, 
but proportional to the second order in the electric field. 
Unlike the conventional (anomalous) Hall effect, 
the nonlinear Hall effect does not necessarily require time-reversal symmetry breaking, 
making it a unique feature of nonlinear responses. 
The physical origins of the nonlinear Hall effect have been well studied, and it is known to arise from band asymmetry~\cite{Morimoto2018,Okumura2021-vz,Michishita2022-ql}, 
Berry curvature dipole~\cite{Sodemann2015-qw,Ma2019-wh}, and quantum metric~\cite{Resta2011-ec,GYN2014,NWang2023} in crystals. 
In particular, the Berry curvature dipole contribution of the nonlinear Hall effect 
is a counterpart of the anomalous Hall effect as both arising from the momentum space topology.

On the other hand, it is known that the Hall effect is caused by the interaction of conduction electrons 
with magnetic textures having scalar spin chirality in real space, 
which is called topological Hall effect~\cite{Ye1999-dk,Tatara2002-ds,Bruno2004-jh,Neubauer2009-js, 
Nakazawa2014-eu, Denisov2016,Nakazawa2018-xp,Nakazawa2019-sb}. 
The topological Hall conductivity is proportional to the density of a topological invariant called the skyrmion number
in systems without spin-orbit interaction. 
Nonlinear Hall effects arising from magnetic textures have also been studied.
The nonlinear Hall effect in magnetic configurations with the vector spin chirality has been reported~\cite{Ishizuka2020-uu}, 
which have been described by local information of magnetic textures, not by topological invariants. 
While there have been
discussions on nonlinear currents in the presence of different types of three-dimensional (topological) magnetic textures 
such as Hopfions~\cite{Liu2022-ei}, they are predicted by symmetry but not specifically shown, 
and their relationship to topological invariants has not been discussed.

Nonlinear transport properties have previously been investigated as higher-order responses to a single driving force. 
Besides, responses to the product of different driving forces, such as the electric field and temperature gradient, 
have also been studied~\cite{Nakai2019-sk}.
In particular, the nonlinear chiral thermo-electric (NCTE) Hall effect, 
where a current flows in the direction of the cross product of the electric field and temperature gradient~\cite{Hidaka2018-vb,Nakai2019-sk,Toshio2020-ax,Yamaguchi2024-yo}, 
is a unique revelation that is expected to occur only in chiral materials, 
as it differs in symmetry from the nonlinear Hall effect caused by a single driving force. 
Recently, we have developed a microscopic theory for the NCTE Hall effect and found that its origin can be described 
not only by the Berry curvature but also by the contribution of the orbital magnetic moment~\cite{Yamaguchi2024-yo}.
We further have examined the NCTE Hall effect quantitatively in chiral tellurium and cobalt monosilicide, 
unveiling the importance of their momentum-space properties~\cite{NYY2024,NYY2024cosi}. 
On the other hand, no research has yet explored the NCTE Hall effect arising from real-space topology, 
which is expected to be realized in the system with the 3D chiral magnetic textures~\cite{Milde2013-dy,Schutte2014-kn,Kanazawa2016-nz,Kagawa2017-un,Fernandez-Pacheco2017-de,Gobel2020-vs,Fujishiro2020-ih,
Pfleiderer2020-zv,Seki2022-fs}.

In this Letter, we theoretically analyze the NCTE Hall effect in chiral magnetic textures based on microscopic models. 
Using nonequilibrium (Keldysh) Green's functions, we describe the NCTE Hall current
due to localized spins by treating the \textit{s-d} exchange interaction perturbatively. 
First, we consider the discrete distribution of the localized spins and show the minimal spin configuration to exhibit
the NCTE Hall effect.
Next, we consider continuous magnetic textures and show that the NCTE Hall conductivity is proportional to the density of magnetic monopoles 
(monopoles created by effective magnetic fields) 
existing in 3D magnetic textures. 
This can be regarded as an extension of the topological Hall effect to the nonlinear regime. 
Our results pave the way to the electrical detection of the total magnetic monopole number in the system. 



\textit{Model and assumptions.--}
We consider a conduction electron system interacting with localized spins and subject to nonmagnetic impurity scattering, 
which is described by the Hamiltonian
\begin{align}
    \mathcal{H} = \mathcal{H}_{\mathrm{el}} + \mathcal{H}_{sd}.
    \label{eq:Hamiltonian}
\end{align}
$\mathcal{H}_{\mathrm{el}}$ consists of the kinetic term and the impurities,
\begin{align}
    \mathcal{H}_{\mathrm{el}} = \int \mathrm{d} \bm{r} c^{\dagger}(\bm{r}) \left[ - \frac{1}{2m} \nabla^{2} + V_{\mathrm{imp}}(\bm{r}) \right] c(\bm{r}),
    \label{eq:Hel}
\end{align}
where $m$ is the electron mass, 
$c^{\dagger} = (c^{\dagger}_{\uparrow}, c^{\dagger}_{\downarrow})$ is the creation operator for conduction electrons, 
and $V_{\mathrm{imp}}(\bm{r}) = u \sum_{i} \delta (\bm{r} - \bm{R}_{i})$ is the impurity potential 
at position $\bm{R}_{i}$ with magnitude $u$.

${\cal H}_{sd}$ represents the \textit{s-d} exchange interaction.
For the localized spins, we consider both discrete magnetic textures and continuous magnetic textures.
The \textit{s-d} exchange interaction for discrete magnetic textures $\mathcal{H}_{sd,\mathrm{dis}}$ can be written as
\begin{align}
    \mathcal{H}_{sd,\mathrm{dis}} = \frac{J}{V} \sum_{\bm{k},\bm{k}'} \sum_{j} \bm{S}_{j} \cdot c^{\dagger}_{\bm{k}} \bm{\sigma} c_{\bm{k}'} \mathrm{e}^{i (\bm{k}' - \bm{k}) \cdot \bm{r}_{j}},
    \label{eq:Hsd_dis}
\end{align}
where $J$ is the coupling constant, $V$ is the volume of the system, 
$\bm{S}_{j}$ represents the localized spin at position $\bm{r}_{j}$, 
$c(\boldsymbol{r}) = V^{-1/2} \sum_{\boldsymbol k} e^{i \boldsymbol k \cdot \boldsymbol r} c_{\boldsymbol k}$,
and $\bm{\sigma} = (\sigma^{x}, \sigma^{y}, \sigma^{z})$ are the Pauli matrices. 
Similarly, the \textit{s-d} exchange interaction $\mathcal{H}_{sd,\mathrm{con}}$ for continuous magnetic textures is given by
\begin{align}
    \mathcal{H}_{sd,\mathrm{con}} = M \int \mathrm{d}\bm{r} \bm{n}(\bm{r}) \cdot c^{\dagger}(\bm{r}) \bm{\sigma} c(\bm{r}),
    \label{eq:Hsd_con}
\end{align}
where $\bm{n}(\bm{r})$ is a unit vector representing the direction of the localized spin at position $\bm{r}$, 
and $M$ is the coupling constant.

The unperturbed retarded Green's function for conduction electrons is defined as
\begin{align}
    G^{\mathrm{R}}_{\bm{k}}(\varepsilon) = \frac{1}{\varepsilon - \varepsilon_{\bm{k}} - \Sigma^{\mathrm{R}}(\varepsilon)},
    \label{eq:GR}
\end{align}
where $\varepsilon_{\bm{k}} = k^{2}/2m$, and $\Sigma^{\mathrm{R}}(\varepsilon)$ is the self-energy, 
which can be written as $\Sigma^{\mathrm{R}}(\varepsilon) = - i \pi n_{\mathrm{i}} u^{2} \nu(\varepsilon) \equiv - i \gamma$ 
within the Born approximation for impurities. 
Here, $\nu(\varepsilon)$ is the density of states of the conduction electrons per unit volume, 
and $n_{\mathrm{i}}$ is the concentration of impurities. 
In the following calculations, we treat the \textit{s-d} exchange interaction as a perturbation in the weak-coupling regime, 
i.e., $J/V, M < \gamma$. 
The transport phenomena in the weak coupling regime have been studied for the anomalous Hall effect 
due to spin chirality~\cite{Tatara2002-ds,Nakazawa2014-eu}
and the topological Hall effect~\cite{Denisov2016,Nakazawa2018-xp,Nakazawa2019-sb},
yielding many physically meaningful results. 
Therefore, the same method may elucidate the magnetic textures which leads to the emergence of the NCTE Hall effect. 
For convenience in future discussions, we define the relaxation time $\tau = 1/(2 \gamma)$, 
the mean free path $\ell = v_{\varepsilon} \tau = k_{\varepsilon} \tau / m$, 
and $k_{\varepsilon} = \sqrt{2 m \varepsilon}$.



\begin{figure}[t]
    \begin{center}
        \includegraphics[width=8cm]{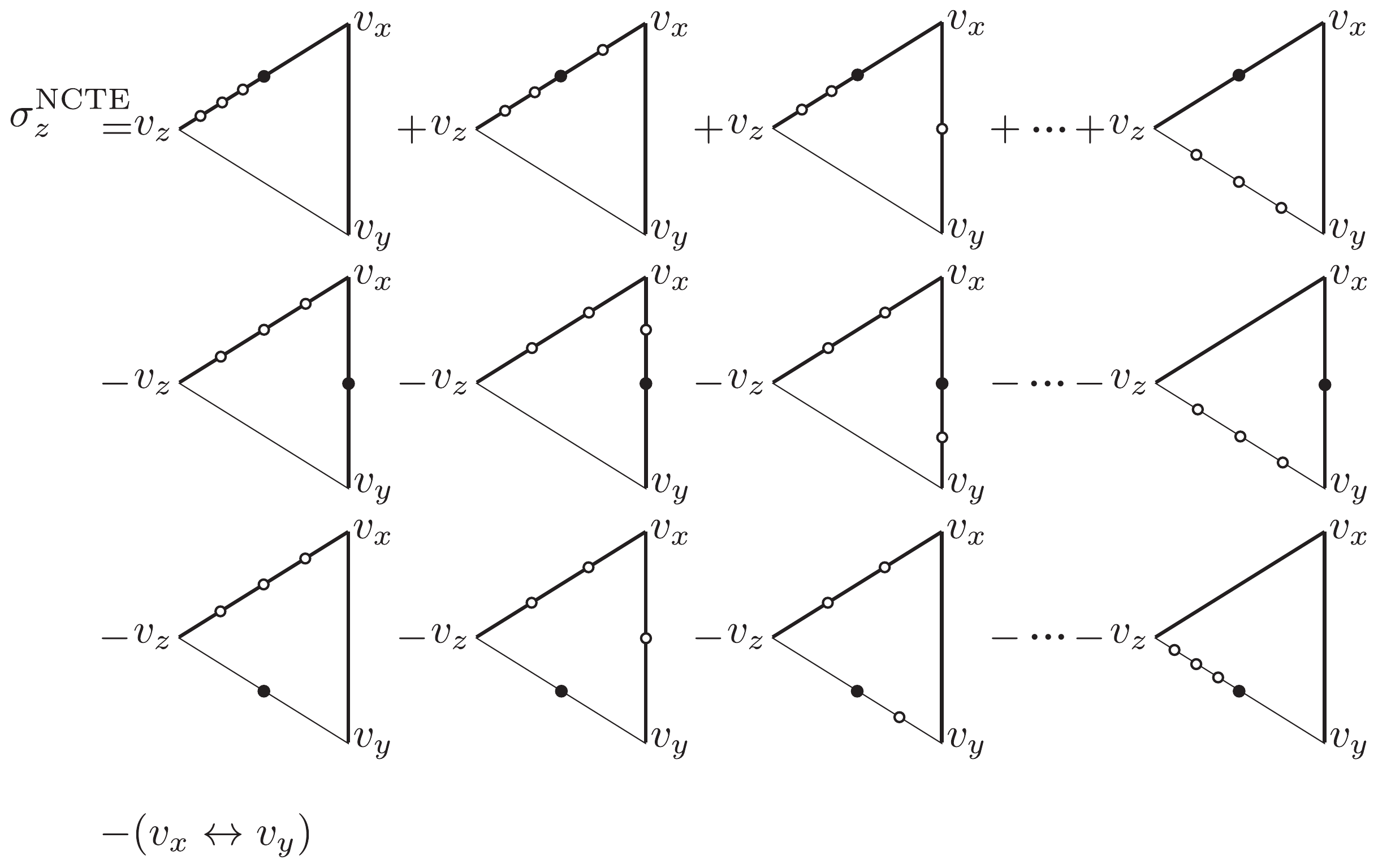}
        \caption{
			Feynman diagrams representing the NCTE Hall conductivity. 
			The bold lines represent the retarded Green's function $G^{\mathrm{R}}$, 
			the thin lines represent the advanced Green's function $G^{\mathrm{A}}$, 
			the closed circles represent the energy derivatives of the retarded (advanced) Green's function 
			$\partial_{\varepsilon} G^{\mathrm{R(A)}} = - (1 \pm i \partial_{\varepsilon}\gamma) (G^{\mathrm{R(A)}})^{2}$, 
			and the open circles represent the perturbation of the \textit{s-d} exchange interaction.
		}
        \label{fig:diagram}
    \end{center}
\end{figure}

\begin{figure}[t]
    \begin{center}
        \includegraphics[width=5cm]{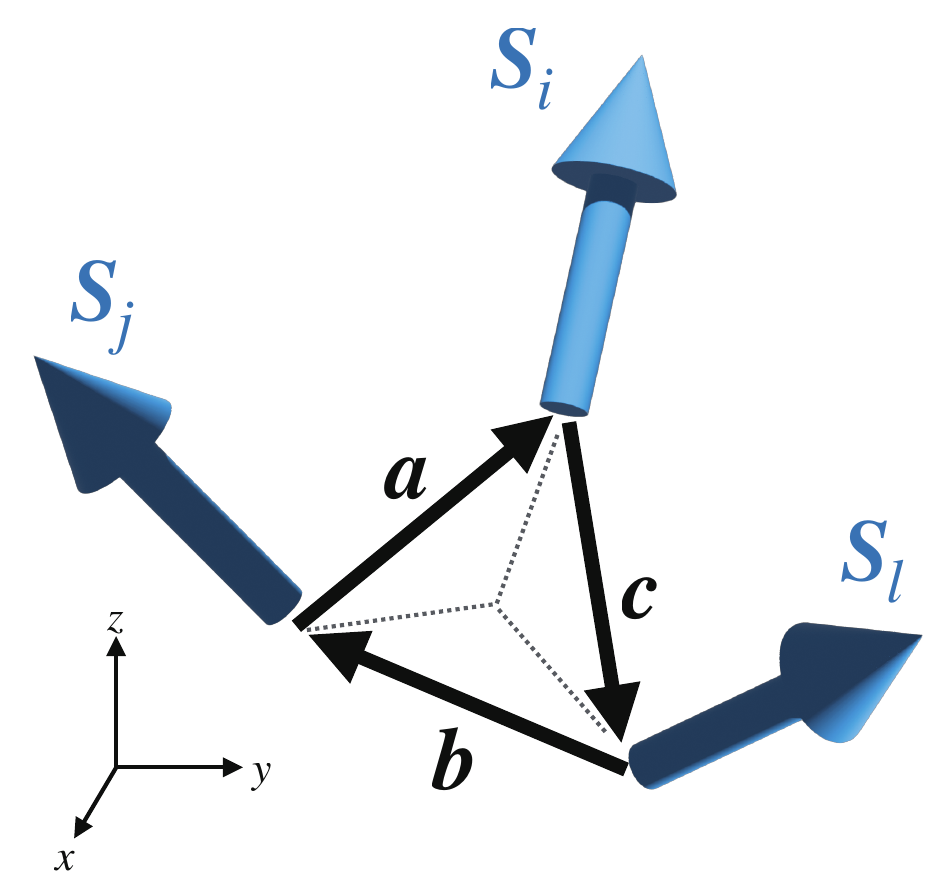}
        \caption{Schematic picture of the arrangement of discrete magnetic textures. 
			$\bm{S}_{i}, \bm{S}_{j}, \bm{S}_{l}$ represent the localized spins at positions $\bm{r}_{i}, \bm{r}_{j}, \bm{r}_{l}$, 
			respectively, and we define
			$\bm{a} = \bm{r}_{i} - \bm{r}_{j}$, $\bm{b} = \bm{r}_{j} - \bm{r}_{l}$, $\bm{c} = \bm{r}_{l} - \bm{r}_{i}$.
		}
        \label{fig:spin_dis}
    \end{center}
\end{figure}



\textit{NCTE Hall effect in the system with discrete magnetization.--}
For simplicity, we consider the case where the electric field and temperature gradient are applied in the $x$-$y$ plane 
and analyze the NCTE Hall current in the $z$ direction without any loss of generality.
The NCTE Hall current can be expressed as
\begin{align}
    j_{z} = \sigma^{\mathrm{NCTE}}_{z} \left[ \bm{E} \times \left( - \frac{\bm{\nabla} T}{T} \right) \right]_{z},
    \label{eq:NCTE}
\end{align}
where $\sigma^{\mathrm{NCTE}}_{z}$ is the NCTE Hall conductivity, 
which can be formulated using the Keldysh Green's function as~\cite{Yamaguchi2024-yo}
\begin{align}
    \sigma^{\mathrm{NCTE}}_{z}
    =&
    - \frac{e^{2}}{4 \pi}
    \int \mathrm{d}\varepsilon \left( - \frac{\partial f}{\partial \varepsilon} \right) (\varepsilon - \mu)
    \int \frac{\mathrm{d}\bm{k}}{(2 \pi)^3}
    \notag \\
    & \times
    \Im
    \left\{ 
        \mathrm{tr}
        \left[ 
            v_{z} \mathcal{G}^{\mathrm{R}} v_{x} (\partial_{\varepsilon} \mathcal{G}^{\mathrm{R}}) v_{y} \mathcal{G}^{\mathrm{R}}
        \right.
    \right.
    \notag \\
    &
    \left.
        \left.
            +
            v_{z} (\partial_{\varepsilon} \mathcal{G}^{\mathrm{R}}) v_{x} \mathcal{G}^{\mathrm{R}} v_{y} \mathcal{G}^{\mathrm{A}}
            -
            v_{z} \mathcal{G}^{\mathrm{R}} v_{x} (\partial_{\varepsilon} \mathcal{G}^{\mathrm{R}}) v_{y} \mathcal{G}^{\mathrm{A}}
        \right.
    \right.
    \notag \\
    &
    \left.
        \left.
            -
            v_{z} \mathcal{G}^{\mathrm{R}} v_{x} \mathcal{G}^{\mathrm{R}} v_{y} (\partial_{\varepsilon} \mathcal{G}^{\mathrm{A}})
        \right]
        - (v_{x} \leftrightarrow v_{y})
    \right\},
    \label{eq:sigmaNCTE}
\end{align}
where $e < 0$ is the electron charge, $v_{i} (i = x,y,z)$ are the velocities, 
and $\mathcal{G}^{\mathrm{R(A)}} = [\varepsilon - \mathcal{H} \pm i 0]^{-1}$ are the full retarded (advanced) Green's functions. 
By treating the $s$-$d$ exchange interaction as a perturbation in Eq.~(\ref{eq:sigmaNCTE}), 
we investigate the NCTE Hall effect due to localized spins. 
First, we perform calculations for the case of discrete magnetization. 
It can be easily confirmed that the first and second orders of the $s$-$d$ exchange interaction vanish, 
thus we focus on the third-order contribution. 
The corresponding Feynman diagrams are shown in Fig.~\ref{fig:diagram}. 
The detailed calculations are given in the Supplemental Material~\cite{supplement}, and the results can be written as
\begin{align}
    \sigma^{\mathrm{NCTE}}_{z}
    =&
    - \int \mathrm{d} \varepsilon \left( - \frac{\partial f}{\partial \varepsilon} \right) (\varepsilon - \mu)
    C(\varepsilon)
    \notag \\
    & \times
    \frac{1}{V}
    \sum_{i,j,l}
    \bm{S}_{i} \cdot \left( \bm{S}_{j} \times \bm{S}_{l} \right)
    \frac{\left( \bm{a} \times \bm{b} \right)_{z} c_{z}}{c^{2}}
    \notag \\
    & \times
    \left[ 
        \frac{I'(a) I(b) I''(c)}{a} + \frac{I(a) I'(b) I''(c)}{b}
    \right],
    \label{NCTE_dis}
\end{align}
where $f(\varepsilon) = 1/(1 + \exp [(\varepsilon - \mu)/k_{\mathrm{B}}T])$ is the Fermi distribution function, 
$\bm{a} = \bm{r}_{i} - \bm{r}_{j}$, $\bm{b} = \bm{r}_{j} - \bm{r}_{l}$, $\bm{c} = \bm{r}_{l} - \bm{r}_{i}$, 
$a = |\bm{a}|$, $b  = |\bm{b}|$, $c = |\bm{c}|$ (see Fig.~\ref{fig:spin_dis}), 
and $I(r) = (\sin k_{\varepsilon} r / k_{\varepsilon} r) \exp [- r/2 \ell]$, 
$I'(r) = \partial_{r} I(r)$, and $I''(r) = \partial_{r}^{2}I(r) - (1/r) \partial_{r} I(r)$. 
The coefficient 
$C(\varepsilon) = 2 e^{2} \pi^{2} (J \nu)^{3} \tau^{4}/m^3$ is also defined. 
Because of the factor $I(r)$ and its derivatives, the effect will be substantial when the distances 
between the localized spins are shorter than the mean free path, $a,b,c < \ell$~\cite{Tatara2002-ds,Nakazawa2014-eu}.

The spin configuration to exhibit the NCTE Hall effect is summarized as follows:
\begin{enumerate}
    \item The Noncoplanar configuration of three localized spins in spin space: 
		$\bm{S}_{i} \cdot \left( \bm{S}_{j} \times \bm{S}_{l} \right) \neq 0$.
    \item The spatial arrangement of the three localized spins to form the 
    finite area in the $xy$-plane: $\left( \bm{a} \times \bm{b} \right)_{z} \neq 0$.
    \item The spin distribution in $z$ direction: $c_{z} \neq 0$.
\end{enumerate}

To exhibit the finite scalar spin chirality leading to the anomalous Hall effect, conditions 1 and 2 are only required~\cite{Tatara2002-ds},
whereas the NCTE Hall effect additionally requires condition 3. 
This means that the triangle formed by the three spins should not lie in the $xy$ plane and, from condition 2, 
cannot be perpendicular to the $xy$ plane either to exhibit the NCTE Hall effect. 
In other words, the NCTE Hall effect induced by the localized spins
is a unique transport phenomenon specific to three-dimensional magnetic textures.
From the factor $(\bm{a} \times \bm{b})_{z}$, the sign changes with the spin configuration around the $z$ axis, 
similar to the spin chirality-induced anomalous Hall effect. 
Additionally, it depends on the sign of $c_{z}$, indicating that the system needs to be chiral, 
with the sign changing for \lq\lq right-handed" or \lq\lq left-handed" chiralities. 



\begin{figure}[t]
    \begin{center}
        \includegraphics[width=8cm]{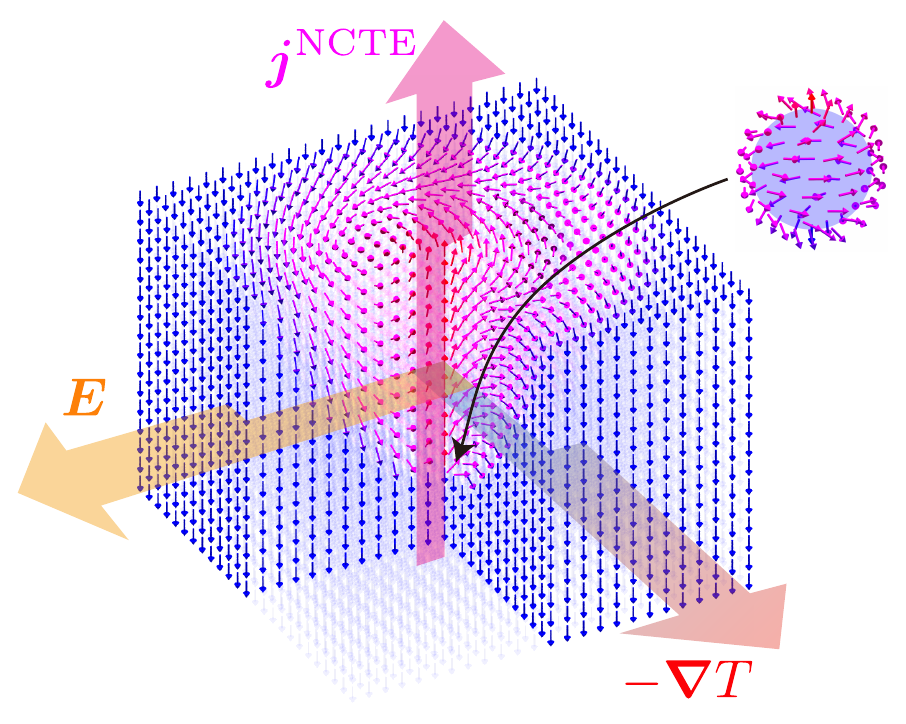}
        \caption{
            Schematic picture of the configuration causing the NCTE Hall effect in the case of continuous magnetization.
            Magnetic monopoles exist at points where skyrmions are generated (annihilated),
            and the NCTE Hall current $\bm{j}^{\mathrm{NCTE}}$ flows proportional to the density of magnetic monopoles.
        }
        \label{fig:spin_con}
    \end{center}
\end{figure}

\textit{NCTE Hall effect in the system with continuous magnetization.--}
We also calculate the NCTE Hall conductivity in the case of continuous magnetization. 
Similar to the discrete magnetization case, we calculate up to the third order in the $s$-$d$ exchange interaction. 
We additionally assume that the spatial variation of the localized spins is smooth compared to any length scale of the electron system,
and can be expanded in terms of its wavenumber. 
Since the first and second orders of the wavenumber integrals are odd functions and thus zero, 
we focus on the third order, which corresponds to the third-order spatial derivative of ${\bm n}({\bm r})$. 
The detailed calculations are given in the Supplemental Material~\cite{supplement}, and the NCTE Hall conductivity is obtained as
\begin{align}
    \sigma^{\mathrm{NCTE}}_{z}
    =&
    \int \mathrm{d} \varepsilon
    \left( - \frac{\partial f}{\partial \varepsilon} \right) (\varepsilon - \mu)
    D(\varepsilon)
    \notag \\
    & \times
    \int \frac{\mathrm{d} \bm{r}}{V}
    \partial_{z} \left[ \bm{n} \cdot \left( \partial_{x} \bm{n} \times \partial_{y} \bm{n} \right) \right],
    \label{eq:NCTE_ave}
\end{align}
where the coefficient is given by 
$D(\varepsilon) = 20 e^{2} \varepsilon^{2} \nu M^{3} \tau^{8}/m^{3}$. 
Namely, the NCTE Hall conductivity can be written as a surface term in the $z$ direction,
the integral in the $xy$ plane becomes the skyrmion number (winding number).
Writing $\partial_{x} \bm{n} \times \partial_{y} \bm{n} = b(\bm{r}) \bm{n}$, 
the skyrmion number in the $xy$ plane at a given $z$ is
$N(z) 
= (4 \pi)^{-1} \iint \mathrm{d}x \mathrm{d}y \bm{n} \cdot (\partial_{x} \bm{n} \times \partial_{y} \bm{n})
= (4 \pi)^{-1} \iint \mathrm{d}x \mathrm{d}y b(\bm{r})
\in \mathbb{Z}$,
then we obtain
\begin{align}
    &\frac{1}{4 \pi} 
    \int \mathrm{d} \bm{r} \partial_{z} \left[ \bm{n} \cdot (\partial_{x} \bm{n} \times \partial_{y} \bm{n}) \right]
    =
    \frac{1}{4 \pi} \int \mathrm{d} \bm{r} \partial_{z} b(\bm{r})
    \notag \\
    &=
    \int_{z_{\mathrm{b}}}^{z_{\mathrm{t}}} \mathrm{d} z \partial_{z} N(z)
    =
    N(z_{\mathrm{t}}) - N(z_{b})
    \in \mathbb{Z},
    \label{eq:integer}
\end{align}
indicating that it turns out to be integer. 
Here $z_{\mathrm{t}}$ and $z_{\mathrm{b}}$ are the coordinates of the top and bottom surfaces in the $z$ direction. 

The term $N(z_{t}) - N(z_{b})$ in Eq.~(\ref{eq:integer}) represents the difference in skyrmion numbers 
between the top and bottom surfaces in the $z$ direction. 
This happens when, for instance,
the creation/annihilation or fusion/fission processes of the skyrmion string occurs, 
which accompany the magnetic monopoles (Bloch points)~\cite{Milde2013-dy,Schutte2014-kn,Kanazawa2016-nz,Kagawa2017-un,Fujishiro2020-ih,
Pfleiderer2020-zv,Seki2022-fs}, depicted in Fig.~\ref{fig:spin_con}.
Therefore, the averaged NCTE Hall current density can be written as
\begin{align}
    j^{\mathrm{NCTE}}_{z}
    =&
    4 \pi
    \mathcal{N}_{\mathrm{m}}
    \left[ \bm{E} \times \left( - \frac{\bm{\nabla}T}{T} \right) \right]_{z}
    \notag \\
    & \times
    \int \mathrm{d} \varepsilon
    \left( - \frac{\partial f}{\partial \varepsilon} \right) (\varepsilon - \mu)
    D(\varepsilon),
    \label{eq:NCTE_ave_integer}
\end{align}
where $\mathcal{N}_{\mathrm{m}}$ is a monopole density.
In the case of the 2D systems, the Hall effect proportional to the skyrmion number 
(without requiring spin-orbit interaction) is called the topological Hall effect. 
Similarly, since the magnetic monopole number is another kind of topological number,
this effect can be regarded as an extension of the topological Hall effect to a nonlinear response in 3D magnetic texture. 
Thus we can term this phenomenon as \lq\lq topological nonlinear Hall effect."



\textit{Discussion.--}
The NCTE Hall effect in the case of discrete magnetization is expected to occur in spin glasses, 
as it shares similar properties with the anomalous Hall effect (linear in electric field) induced by spin chirality. 
However, it is important to note that the NCTE Hall effect can occur in time-reversal symmetric systems, 
so it can arise even in spin glasses without net magnetization, in contrast to the chirality-driven Hall effect~\cite{Tatara2002-ds}.
Additionally, the NCTE Hall effect may occur in chiral magnets due to the discrete spins on magnetic elements. 
For example, applying a magnetic field to chiral magnets such as MnSi~\cite{Tonomura2012-wm,Yu2013-bv,Fujishiro2020-ih,Tokura2021-qp}
or MnAu$_{2}$~\cite{Glasbrenner2014-ef,Glasbrenner2016-ga,Masuda2024-ra} induces spin configuration such that
$[\bm{S}_{i} \cdot (\bm{S}_{j} \times \bm{S}_{l})] ({\bm a} \times {\bm b})_{z} c_{z} \neq 0$, 
creating a chiral atomic arrangement that satisfies the conditions for the emergence of NCTE Hall effect. 

When considering continuous magnetization, several points need attention. 
In this letter, we found that the NCTE Hall conductivity is proportional to the density of magnetic monopoles, 
including their \lq\lq sign." 
For example, if skyrmions appear along the positive $z$-axis (as shown in Fig.~\ref{fig:spin_con}), 
they have a charge of $+1$, while disappearing skyrmions have a charge of $-1$. 
When both are present in equal numbers, the net charge is zero 
(e.g., magnetic hedgehog lattice~\cite{Fujishiro2020-ih,OHKY2020,SOKY2021,SOKY2022}).
However, as we mentioned before, in the presence of the creation/annihilation or fusion/fission processes of the skyrmion tubes, 
the total charge can be finite and detected using the NCTE Hall effect. 
It will be interesting to compare the results with the real-space visualization of the magnetic texture~\cite{Yu2013-bv,Yu2022} 
which can directly count the number of skyrmion at the top and bottom surfaces. 

We comment on the relaxation time dependence obtained from the calculations and its relation to the vertex correction. 
In this study, the NCTE Hall conductivity is proportional to $\tau^4$ and $\tau^8$ in the case of discrete and continuous magnetization, respectively. 
Such a large power of $\tau$ appears in the nonlinear Hall conductivity in the system with vector spin chirality ($\tau^{4}$)~\cite{Ishizuka2020-uu} 
and even in the linear response coefficient, 
since the number of Green's function in the diagram increases in case of simple perturbative treatment of $s$-$d$ interactions~\cite{Nakazawa2019-sb,Yamaguchi2021}. 
In the weak coupling regime, the $\tau$ dependence can be suppressed by considering vertex corrections due to impurity scattering~\cite{Nakazawa2019-sb}. 
Another interesting effect of the vertex correction will be the spin diffusion of electrons. 
In this case, the electrons acquire the information of farther spins~\cite{HLN1980,Nakazawa2014-eu}, 
which hinder the topological description~\cite{Nakazawa2018-xp,Nakazawa2019-sb}. 
However, we can expect the finite NCTE Hall conductivity even in the case of same number of opposite monopole charges that we mentioned before. 
For instance, if the spatial modulation of localized spins around monopole charge +1 is fast while that of $-1$ is slow, 
the system can be chiral leading to \lq\lq nonlocal" NCTE Hall effect.
These considerations are crucial for quantitative evaluation, 
but they are beyond the scope of this letter and are pointed out as future challenges.



\textit{Conclusion.--}
In this letter, we have investigated the NCTE Hall effect arising from magnetic textures 
in both discrete and continuous magnetization cases. 
The calculations for discrete magnetization clarified the conditions for the emergence of the NCTE Hall effect, 
revealing the necessity of not only scalar spin chirality but also a three-dimensional spatial structure. 
From these results, we discussed the possibility of observing the effect in spin glasses and chiral magnets. 
Furthermore, the analysis for continuous magnetization showed that 
the NCTE Hall effect is proportional to the density of magnetic monopoles. 
This finding represents the first extension of the topological Hall effect to a nonlinear regime, 
and we also discussed potential systems where this effect could be experimentally observed. 
This study focuses on phenomena emerging under three-dimensional magnetic textures rather than two-dimensional ones, 
highlighting the importance of considering nonlinear responses beyond conventional linear responses, 
and even more so, the significance of combining multiple driving forces. 
This work opens up new directions in condensed matter physics, particularly in the field of spintronics and quantum transport.




\begin{acknowledgments}
 We acknowledge H. Kohno, H. Ishizuka, Y. Nishida, T. Yamazaki and T. Hioki for fruitiful discussion.
 A.Y. is supported by JSPS KAKENHI (Grant No. JP20K03835). 
 K.N. is supported by JSPS KAKENHI (Grant No. JP21K13875). 
 T.Y. is supported by JSPS KAKENHI (Grant No. JP21K14526).
\end{acknowledgments}

\appendix

\onecolumngrid

\section{Detail calculation of NCTE Hall conductivity in the system with discrete magnetization}

In this section, we show the calculation of NCTE Hall conductivity in the system with discrete magnetization.
From the Feynman diagrams shown in Fig. 1 in main text, we can express the NCTE Hall current as
\begin{align}
	j^{\mathrm{NCTE}}_{z}
	=
	\frac{e^{2}}{4 \pi V} \left( \frac{J}{V} \right)^{3}
	\int \mathrm{d} \varepsilon \left( - \frac{\partial f}{\partial \varepsilon} \right) \varepsilon
	\left[ 
		\eta^{\mathrm{I}} - \eta^{\mathrm{II}} - \eta^{\mathrm{III}}
	\right]
	\left[ 
		\bm{E} \times \left( - \frac{\bm{\nabla} T}{T} \right)
	\right]_{z},
\end{align}
where
\begin{align}
	\eta^{\mathrm{I}}
	=&
	\Im
	\Biggl[
		(1 + i \partial_{\varepsilon} \gamma)
		\sum_{i,j,l} \bm{S}_{i} \cdot (\bm{S}_{j} \times \bm{S}_{l})
		\sum_{\bm{k}_{1},\bm{k}_{2},\bm{k}_{3}}
		\mathrm{e}^{i (\bm{k}_{2} - \bm{k}_{1}) \cdot \bm{r}_{i}}
		\mathrm{e}^{i (\bm{k}_{3} - \bm{k}_{2}) \cdot \bm{r}_{j}}
		\mathrm{e}^{i (\bm{k}_{1} - \bm{k}_{3}) \cdot \bm{r}_{l}}
	\notag \\
	& \times
		\left\{ 
			\left[ 
				v_{x} v_{y} v_{z}
				(G^{\mathrm{R}})^{4} G^{\mathrm{A}}
			\right]_{\bm{k}_{1}}
			\left[ G^{\mathrm{R}} \right]_{\bm{k}_{2}}
			\left[ G^{\mathrm{R}} \right]_{\bm{k}_{3}}
			+
			\left[ 
				v_{x} v_{z} (G^{\mathrm{R}})^{3} G^{\mathrm{A}}
			\right]_{\bm{k}_{1}}
			\left[ G^{\mathrm{R}} \right]_{\bm{k}_{2}}
			\left[ v_{y} G^{\mathrm{R}} G^{\mathrm{A}} \right]_{\bm{k}_{3}}
		\right.
	\notag \\
	& \quad
		\left.
			+
			\left[ 
				v_{x} v_{z} (G^{\mathrm{R}})^{3} G^{\mathrm{A}}
			\right]_{\bm{k}_{1}}
			\left[ v_{y} G^{\mathrm{R}} G^{\mathrm{A}} \right]_{\bm{k}_{2}}
			\left[ G^{\mathrm{A}} \right]_{\bm{k}_{3}}
		\right.
	\notag \\
	& \quad
		\left.
			+
			\left[ 
				v_{y} v_{z} (G^{\mathrm{R}})^{3} G^{\mathrm{A}}
			\right]_{\bm{k}_{1}}
			\left[ G^{\mathrm{R}} \right]_{\bm{k}_{2}}
			\left[ v_{x} (G^{\mathrm{R}})^{2} \right]_{\bm{k}_{3}}
			+
			\left[ v_{y} v_{z} (G^{\mathrm{R}})^{3} G^{\mathrm{A}} \right]_{\bm{k}_{1}}
			\left[ v_{x} (G^{\mathrm{R}})^{2} \right]_{\bm{k}_{2}}
			\left[ G^{\mathrm{R}} \right]_{\bm{k}_{3}}
		\right.
	\notag \\
	& \quad
		\left.
			+
			\left[ v_{z} (G^{\mathrm{R}})^{2} G^{\mathrm{A}} \right]_{\bm{k}_{1}}
			\left[ v_{x} (G^{\mathrm{R}})^{2} \right]_{\bm{k}_{2}}
			\left[ v_{y} G^{\mathrm{R}} G^{\mathrm{A}} \right]_{\bm{k}_{3}}
		\right\}
		-
		\left\{ x \leftrightarrow y \right\}
	\Biggr],
	\\
	\eta^{\mathrm{II}}
	=&
	\Im
	\Biggl[
		(1 + i \partial_{\varepsilon} \gamma)
		\sum_{i,j,l} \bm{S}_{i} \cdot (\bm{S}_{j} \times \bm{S}_{l})
		\sum_{\bm{k}_{1},\bm{k}_{2},\bm{k}_{3}}
		\mathrm{e}^{i (\bm{k}_{2} - \bm{k}_{1}) \cdot \bm{r}_{i}}
		\mathrm{e}^{i (\bm{k}_{3} - \bm{k}_{2}) \cdot \bm{r}_{j}}
		\mathrm{e}^{i (\bm{k}_{1} - \bm{k}_{3}) \cdot \bm{r}_{l}}
	\notag \\
	& \times
		\left\{ 
			\left[ v_{x} v_{y} v_{z} (G^{\mathrm{R}})^{4} G^{\mathrm{A}} \right]_{\bm{k}_{1}}
			\left[ G^{\mathrm{R}} \right]_{\bm{k}_{2}}
			\left[ G^{\mathrm{R}} \right]_{\bm{k}_{3}}
			+
			\left[ v_{x} v_{z} (G^{\mathrm{R}})^{2} G^{\mathrm{A}} \right]_{\bm{k}_{1}}
			\left[ G^{\mathrm{R}} \right]_{\bm{k}_{2}}
			\left[ v_{y} (G^{\mathrm{R}})^{2} G^{\mathrm{A}} \right]_{\bm{k}_{3}}
		\right.
	\notag \\
	& \quad
		\left.
			+
			\left[ v_{x} v_{z} (G^{\mathrm{R}})^{2} G^{\mathrm{A}} \right]_{\bm{k}_{1}}
			\left[ v_{y} (G^{\mathrm{R}})^{2} G^{\mathrm{A}} \right]_{\bm{k}_{2}}
			\left[ G^{\mathrm{A}} \right]_{\bm{k}_{3}}
			+
			\left[ v_{x} v_{y} v_{z} (G^{\mathrm{R}})^{4} G^{\mathrm{A}} \right]_{\bm{k}_{1}}
			\left[ G^{\mathrm{R}} \right]_{\bm{k}_{2}}
			\left[ G^{\mathrm{R}} \right]_{\bm{k}_{3}}
		\right.
	\notag \\
	& \quad
		\left.
			+
			\left[ v_{x} v_{z} (G^{\mathrm{R}})^{3} G^{\mathrm{A}} \right]_{\bm{k}_{1}}
			\left[ G^{\mathrm{R}} \right]_{\bm{k}_{2}}
			\left[ v_{y} G^{\mathrm{R}} G^{\mathrm{A}} \right]_{\bm{k}_{3}}
			+
			\left[ v_{x} v_{z} (G^{\mathrm{R}})^{3} G^{\mathrm{A}} \right]_{\bm{k}_{1}}
			\left[ v_{y} G^{\mathrm{R}} G^{\mathrm{A}} \right]_{\bm{k}_{2}}
			\left[ G^{\mathrm{A}} \right]_{\bm{k}_{3}}
		\right.
	\notag \\
	& \quad
		\left.
			+
			\left[ v_{y} v_{z} (G^{\mathrm{R}})^{3} G^{\mathrm{A}} \right]_{\bm{k}_{1}}
			\left[ G^{\mathrm{R}} \right]_{\bm{k}_{2}}
			\left[ v_{x} (G^{\mathrm{R}})^{2} \right]_{\bm{k}_{3}}
			+
			\left[ v_{y} v_{z} (G^{\mathrm{R}})^{3} G^{\mathrm{A}} \right]_{\bm{k}_{1}}
			\left[ v_{x} (G^{\mathrm{R}})^{2} \right]_{\bm{k}_{2}}
			\left[ G^{\mathrm{R}} \right]_{\bm{k}_{3}}
		\right.
	\notag \\
	& \quad
		\left.
			+
			\left[ v_{z} G^{\mathrm{R}} G^{\mathrm{A}} \right]_{\bm{k}_{1}}
			\left[ v_{x} (G^{\mathrm{R}})^{2} \right]_{\bm{k}_{2}}
			\left[ v_{y} (G^{\mathrm{R}})^{2} G^{\mathrm{A}} \right]_{\bm{k}_{3}}
			+
			\left[ v_{x} v_{y} v_{z} (G^{\mathrm{R}})^{3} G^{\mathrm{A}} \right]_{\bm{k}_{1}}
			\left[ G^{\mathrm{R}} \right]_{\bm{k}_{2}}
			\left[ (G^{\mathrm{R}})^{2} \right]_{\bm{k}_{3}}
		\right.
	\notag \\
	& \quad
		\left.
			+
			\left[ v_{x} v_{y} v_{z} (G^{\mathrm{R}})^{3} G^{\mathrm{A}} \right]_{\bm{k}_{1}}
			\left[ (G^{\mathrm{R}})^{2} \right]_{\bm{k}_{2}}
			\left[ G^{\mathrm{R}} \right]_{\bm{k}_{3}}
			+
			\left[ v_{x} v_{z} (G^{\mathrm{R}})^{2} G^{\mathrm{A}} \right]_{\bm{k}_{1}}
			\left[ (G^{\mathrm{R}})^{2} \right]_{\bm{k}_{2}}
			\left[ v_{y} G^{\mathrm{R}} G^{\mathrm{A}} \right]_{\bm{k}_{3}}
		\right\}
		-
		\left\{ x \leftrightarrow y \right\}
	\Biggr],
	\\
	\eta^{\mathrm{III}}
	=&
	\Im
	\Biggl[
		(1 - i \partial_{\varepsilon} \gamma)
		\sum_{i,j,l} \bm{S}_{i} \cdot (\bm{S}_{j} \times \bm{S}_{l})
		\sum_{\bm{k}_{1},\bm{k}_{2},\bm{k}_{3}}
		\mathrm{e}^{i (\bm{k}_{2} - \bm{k}_{1}) \cdot \bm{r}_{i}}
		\mathrm{e}^{i (\bm{k}_{3} - \bm{k}_{2}) \cdot \bm{r}_{j}}
		\mathrm{e}^{i (\bm{k}_{1} - \bm{k}_{3}) \cdot \bm{r}_{l}}
	\notag \\
	& \times
		\left\{ 
			\left[ v_{x} v_{y} v_{z} (G^{\mathrm{R}})^{3} (G^{\mathrm{A}})^{2} \right]_{\bm{k}_{1}}
			\left[ G^{\mathrm{R}} \right]_{\bm{k}_{2}}
			\left[ G^{\mathrm{R}} \right]_{\bm{k}_{3}}
			+
			\left[ v_{x} v_{z} (G^{\mathrm{R}})^{2} G^{\mathrm{A}} \right]_{\bm{k}_{1}}
			\left[ v_{y} G^{\mathrm{R}} (G^{\mathrm{A}})^{2} \right]_{\bm{k}_{2}}
			\left[ G^{\mathrm{A}} \right]_{\bm{k}_{3}}
		\right.
	\notag \\
	& \quad
		\left.
			+
			\left[ v_{x} v_{z} (G^{\mathrm{R}})^{2} (G^{\mathrm{A}})^{2} \right]_{\bm{k}_{1}}
			\left[ G^{\mathrm{R}} \right]_{\bm{k}_{2}}
			\left[ v_{y} G^{\mathrm{R}} G^{\mathrm{A}} \right]_{\bm{k}_{3}}
			+
			\left[ v_{x} v_{z} (G^{\mathrm{R}})^{2} (G^{\mathrm{A}})^{2} \right]_{\bm{k}_{1}}
			\left[ v_{y} G^{\mathrm{R}} G^{\mathrm{A}} \right]_{\bm{k}_{2}}
			\left[ G^{\mathrm{A}} \right]_{\bm{k}_{3}}
		\right.
	\notag \\
	& \quad
		\left.
			+
			\left[ v_{x} v_{z} (G^{\mathrm{R}})^{2} G^{\mathrm{A}} \right]_{\bm{k}_{1}}
			\left[ G^{\mathrm{R}} \right]_{\bm{k}_{2}}
			\left[ v_{y} G^{\mathrm{R}} (G^{\mathrm{A}})^{2} \right]_{\bm{k}_{3}}
		\right.
	\notag \\
	& \quad
		\left.
			+
			\left[ v_{x} v_{z} (G^{\mathrm{R}})^{2} G^{\mathrm{A}} \right]_{\bm{k}_{1}}
			\left[ v_{y} G^{\mathrm{R}} G^{\mathrm{A}} \right]_{\bm{k}_{2}}
			\left[ (G^{\mathrm{A}})^{2} \right]_{\bm{k}_{3}}
		\right\}
		-
		\left\{ x \leftrightarrow y \right\}
	\Biggr].
\end{align}
Here we use the notation
$\left[ (G^{\mathrm{R}})^{n} (G^{\mathrm{A}})^{m} \right]_{\bm{k}_{i}} 
= \left( G^{\mathrm{R}}_{\bm{k}_{i}}(\varepsilon) \right)^{n} \left( G^{\mathrm{A}}_{\bm{k}_{i}}(\varepsilon) \right)^{m}$ 
for simplicity.
To perform wavenumber integrals, we consider the integral shown below,
\begin{align}
    \frac{1}{V}
	\sum_{\bm{k}}
	G^{\mathrm{R}}_{\bm{k}} G^{\mathrm{A}}_{\bm{k}} \mathrm{e}^{i \bm{k} \cdot \bm{r}}
	=&
	2 \pi
	\int \mathrm{d} \theta \sin \theta
	\int \mathrm{d} k \ k^{2}
	G^{\mathrm{R}}_{k} G^{\mathrm{A}}_{k} \mathrm{e}^{i kr \cos \theta}
	=
	2 \pi
	\int \mathrm{d} k \ k^{2}
	\frac{\mathrm{e}^{i k r} - \mathrm{e}^{-ikr}}{i k r}
	G^{\mathrm{R}}_{k} G^{\mathrm{A}}_{k}
	\notag \\
	=&
	\int \mathrm{d} \xi \nu(\xi)
	\frac{\mathrm{e}^{ik_{\xi} r} - \mathrm{e}^{- i k_{\xi} r}}{2 i k_{\xi} r}
	\frac{1}{(\xi - \varepsilon - i \gamma)(\xi - \varepsilon + i \gamma)}
	\notag \\
	=&
	2 i \pi
	\left[ 
		\frac{\mathrm{e}^{i k_{\varepsilon + i \gamma} r}}{2 i k_{\varepsilon + i \gamma} r} 
		\frac{\nu(\varepsilon + i \gamma)}{2 i \gamma}
		+
		\frac{\mathrm{e}^{- i k_{\varepsilon - i \gamma} r}}{- 2 i k_{\varepsilon - i \gamma} r} 
		\frac{-\nu(\varepsilon - i \gamma)}{(- 2 i \gamma)}
	\right]
	\notag \\
	\simeq &
	\frac{\pi \nu(\varepsilon)}{\gamma}
	\mathrm{e}^{-\frac{r}{2 v_{\varepsilon} \tau}}
	\frac{1}{k_{\varepsilon} r}
	\left[ 
		\sin k_{\varepsilon} r
		+
		\left\{ \gamma (\partial_{\varepsilon} \nu) - \frac{1}{2 k_{\varepsilon} v_{\varepsilon} \tau} \right\}
		\cos k_{\varepsilon} r
	\right],
\end{align}
where $\nu(\xi)$ is the density of states, and $k_{\xi} = \sqrt{2 m \xi}$.
In last line, we use $\gamma \ll \varepsilon_{\mathrm{F}}$ and retain 
up to the second lowest order in $\gamma/\varepsilon_{\mathrm{F}}$.
Next, we consider the case including the velocities.
The velocity is defined in our model as $v_{i} = k_{i}/m$,
then we obtain
\begin{align}
    &\frac{1}{V}
	\sum_{\bm{k}}
	v_{j} 
	(G^{\mathrm{R}})^{n} (G^{\mathrm{A}})^{\ell} \mathrm{e}^{i \bm{k} \cdot \bm{r}}
	=
	\frac{1}{i m} \frac{\partial}{\partial r_{j}}
	\sum_{\bm{k}}
	(G^{\mathrm{R}})^{n} (G^{\mathrm{A}})^{\ell} \mathrm{e}^{i \bm{k} \cdot \bm{r}}
	=
	\frac{1}{i m} \frac{x_{j}}{r} \frac{\partial}{\partial r}
	\sum_{\bm{k}}
	(G^{\mathrm{R}})^{n} (G^{\mathrm{A}})^{\ell} \mathrm{e}^{i \bm{k} \cdot \bm{r}},
	\\
    &\frac{1}{V}
	\sum_{\bm{k}}
	v_{j_{1}} v_{j_{2}}
	(G^{\mathrm{R}})^{n} (G^{\mathrm{A}})^{\ell}
	\mathrm{e}^{i \bm{k}\cdot \bm{r}}
	=
	\left[
		- 
		\frac{\delta_{j_{1},j_{2}}}{m^{2}}
		\frac{1}{r} \frac{\partial}{\partial r}
		- 
		\frac{1}{m^{2}} \frac{x_{j_{1}} x_{j_{2}}}{r^{2}}
		\left\{ 
			\frac{\partial^{2}}{\partial r^{2}} - \frac{1}{r} \frac{\partial}{\partial r}
		\right\}
	\right]
	\sum_{\bm{k}} (G^{\mathrm{R}})^{n} (G^{\mathrm{A}})^{\ell}
	\mathrm{e}^{i \bm{k}\cdot \bm{r}}.
\end{align}
Using the above integrals and the relation,
\begin{align}
	\mathrm{e}^{i (\bm{k}_{2} - \bm{k}_{1}) \cdot \bm{r}_{i}}
	\mathrm{e}^{i (\bm{k}_{3} - \bm{k}_{2}) \cdot \bm{r}_{j}}
	\mathrm{e}^{i (\bm{k}_{1} - \bm{k}_{3}) \cdot \bm{r}_{l}}
	=
	\mathrm{e}^{i \bm{k}_{1} (\bm{r}_{l} - \bm{r}_{i})}
	\mathrm{e}^{i \bm{k}_{2} (\bm{r}_{i} - \bm{r}_{j})}
	\mathrm{e}^{i \bm{k}_{3} (\bm{r}_{j} - \bm{r}_{l})}
	=
	\mathrm{e}^{i \bm{k}_{1} \bm{c}}
	\mathrm{e}^{i \bm{k}_{2} \bm{a}}
	\mathrm{e}^{i \bm{k}_{3} \bm{b}},
\end{align}
where we define 
$\bm{r}_{i} - \bm{r}_{j} = \bm{a}$, $\bm{r}_{j} - \bm{r}_{l} = \bm{b}$, and $\bm{r}_{l} - \bm{r}_{i} = \bm{c}$.
Therefore, we obtain the calculation results,
\begin{align}
	\eta^{\mathrm{I}}
	\simeq & 0,
	\\
	\eta^{\mathrm{II}}
	\simeq &
	\frac{1}{m^{3}}
	\frac{c_{z} (c_{x} b_{y} - c_{y} b_{x})}{b c^{2}} \frac{(\pi \nu)^{3}}{4 \gamma^{4}}
	I(a) I'(b) I''(c)
	- 
	\frac{1}{m^{3}}
	\frac{c_{z} (c_{x} a_{y} - c_{y} c_{x})}{a c^{2}} \frac{(\pi \nu)^{3}}{4 \gamma^{4}}
	I'(a) I(b) I''(c),
	\\
	\eta^{\mathrm{III}}
	\simeq &
	\frac{1}{m^{3}}
	\frac{c_{z} (c_{x} a_{y} - c_{y} c_{x})}{a c^{2}}
	\frac{(\pi \nu)^{3}}{4 \gamma^{4}}
	I'(a) I(b) I''(c)
	-
	\frac{1}{m^{3}}
	\frac{c_{z} (c_{x} b_{y} - c_{y} c_{x})}{b c^{2}}
	\frac{(\pi \nu)^{3}}{2 \gamma^{4}}
	I(a) I'(b) I''(c)
	\notag \\
	&
	+ 
	\frac{1}{m^{3}}
	\frac{c_{z} (c_{x} a_{y} - c_{y} c_{x})}{a c^{2}}
	\frac{(\pi \nu)^{3}}{2 \gamma^{4}}
	I'(a) I(b) I''(c)
	-
	\frac{1}{m^{3}}
	\frac{c_{z} (c_{x} b_{y} - c_{y} c_{x})}{b c^{2}}
	\frac{(\pi \nu)^{3}}{4 \gamma^{4}}
	I(a) I'(b) I''(c),
\end{align}
where we define
$I'(r) = \partial_{r}I(r)$ and $I''(r)= \partial_{r}^{2} I(r) - (1/r) \partial_{r} I(r)$.
The NCTE Hall conductivity is described as
\begin{align}
	\sigma^{\mathrm{NCTE}}_{z}
	=&
	\frac{e^{2}}{4 \pi} \left( \frac{J}{V} \right)^{3}
	\int \mathrm{d} \varepsilon \left( - \frac{\partial f}{\partial \varepsilon} \right) \varepsilon
	\frac{1}{m^{3}}
	\frac{(\pi \nu)^{3}}{2 \gamma^{4}}
    \notag \\
    & \times
    \frac{1}{V} \sum_{i,j,l}
	\bm{S}_{i} \cdot (\bm{S}_{j} \times \bm{S}_{l})
	\frac{1}{c}
	\left[ 
		\frac{c_{z} (\bm{c} \times \bm{b})_{z}}{bc}
		I(a) I'(b) I''(c)
		-
		\frac{c_{z} (\bm{c} \times \bm{a})_{z}}{ac}
		I'(a) I(b) I''(c)
	\right]
	\notag \\
	=&
	-
	\int \mathrm{d} \varepsilon \left( - \frac{\partial f}{\partial \varepsilon} \right) \varepsilon
	C(\varepsilon)
    \frac{1}{V} \sum_{i,j,l}
	\bm{S}_{i} \cdot (\bm{S}_{j} \times \bm{S}_{l})
	\frac{c_{z} (\bm{a} \times \bm{b})_{z}}{c^{2}}
	\left[ 
		\frac{1}{a} I'(a) I(b) I''(c)
		+
		\frac{1}{b} I(a) I'(b) I''(c)
	\right],
\end{align}
with
$C(\varepsilon) 
= 2 e^{2} \pi^{2} (J \nu)^{3} \tau^{4}/m^3$.

\section{Detail calculation of NCTE Hall conductivity in the system with continuous magnetization}

In this section, we show the detail calculations of NCTE Hall conductivity in the system with continuous magnetization.
From the Feynman diagrams shown in Fig. 1 in main text, and noting the relation of Pauli matrices,
$\sigma^{\alpha} \sigma^{\beta} = \delta_{\alpha \beta} I_{2 \times 2} + i \epsilon_{\alpha \beta \gamma} \sigma^{\gamma}$,
the NCTE Hall current can be written as
\begin{align}
	j^{\mathrm{NCTE}}_{z}
	=&
	\frac{e^{2} M^{3}}{4 \pi}
	\int \mathrm{d} \varepsilon
	\left( - \frac{\partial f}{\partial \varepsilon} \right) (\varepsilon - \mu)
    \int \frac{\mathrm{d}\bm{r}}{V}
	\left[
		\chi^{\mathrm{I}} - \chi^{\mathrm{II}} - \chi^{\mathrm{III}}
	\right]
	\left[ 
		\bm{E} \times \left( - \frac{\bm{\nabla} T}{T} \right)
	\right]_{z},
\end{align}
where
\begin{align}
	\chi^{\mathrm{I}}
	=&
	\Im
	\Biggl[
		2 i
		\left\{ 1 + i \partial_{\varepsilon} \gamma \right\}
		\sum_{\bm{q}_{1},\bm{q}_{2},\bm{q}_{3}} \mathcal{S} (\bm{q}_{1}, \bm{q}_{2},\bm{q}_{3}) 
		\mathrm{e}^{i \bm{q} \cdot \bm{r}}
	\notag \\
	& \quad
		\times
		\sum_{\bm{k}}
		\left\{ 
			v_{z}
			\left( v_{x}^{+ - -} v_{y}^{- - -} - v_{y}^{+ - -} v_{x}^{- - -} \right)
			G^{\mathrm{R}}_{+ + +} G^{\mathrm{R}}_{+ + -} (G^{\mathrm{R}}_{+ - -})^{3} G^{\mathrm{R}}_{- - -}
			G^{\mathrm{A}}_{- - -}
		\right.
	\notag \\
	& \qquad
		\left.
			+
			v_{z}
			\left( v_{x}^{+ - -} v_{y}^{- - -} - v_{y}^{+ - -} v_{x}^{- - -} \right)
			G^{\mathrm{R}}_{+ + +} (G^{\mathrm{R}}_{+ + -})^{2} (G^{\mathrm{R}}_{+ - -})^{2} G^{\mathrm{R}}_{- - -}
			G^{\mathrm{A}}_{- - -}
		\right.
	\notag \\
	& \qquad
		\left.
			+
			v_{z}
			\left( v_{x}^{+ + -} v_{y}^{- - -} - v_{y}^{+ + -} v_{x}^{- - -} \right)
			G^{\mathrm{R}}_{+ + +} (G^{\mathrm{R}}_{+ + -})^{3} G^{\mathrm{R}}_{+ - -} G^{\mathrm{R}}_{- - -}
			G^{\mathrm{A}}_{- - -}
		\right.
	\notag \\
	& \qquad
		\left.
			+
			v_{z}
			\left( v_{x}^{+ + -} v_{y}^{+ - -} - v_{y}^{+ + -} v_{x}^{+ - -} \right)
			G^{\mathrm{R}}_{+ + +} (G^{\mathrm{R}}_{+ + -})^{3} G^{\mathrm{R}}_{+ - -}
			G^{\mathrm{A}}_{+ - -} G^{\mathrm{A}}_{- - -}
		\right.
	\notag \\
	& \qquad
		\left.
			+
			v_{z}
			\left( v_{x}^{+ - -} v_{y}^{- - -} - v_{y}^{+ - -} v_{x}^{- - -} \right)
			(G^{\mathrm{R}}_{+ + +})^{2} G^{\mathrm{R}}_{+ + -} (G^{\mathrm{R}}_{+ - -})^{2} G^{\mathrm{R}}_{- - -}
			G^{\mathrm{A}}_{- - -}
		\right.
	\notag \\
	& \qquad
		\left.
			+
			v_{z}
			\left( v_{x}^{+ + -} v_{y}^{- - -} - v_{y}^{+ + -} v_{x}^{- - -} \right)
			(G^{\mathrm{R}}_{+ + +})^{2} (G^{\mathrm{R}}_{+ + -})^{2} G^{\mathrm{R}}_{+ - -} G^{\mathrm{R}}_{- - -}
			G^{\mathrm{A}}_{- - -}
		\right.
	\notag \\
	& \qquad
		\left.
			+
			v_{z}
			\left( v_{x}^{+ + -} v_{y}^{+ - -} - v_{y}^{+ + -} v_{x}^{+ - -} \right)
			(G^{\mathrm{R}}_{+ + +})^{2} (G^{\mathrm{R}}_{+ + -})^{2} G^{\mathrm{R}}_{+ - -}
			G^{\mathrm{A}}_{+ - -} G^{\mathrm{A}}_{- - -}
		\right.
	\notag \\
	& \qquad
		\left.
			+
			v_{z}
			\left( v_{x}^{+ + +} v_{y}^{- - -} - v_{y}^{+ + +} v_{x}^{- - -} \right)
			(G^{\mathrm{R}}_{+ + +})^{3} G^{\mathrm{R}}_{+ + -} G^{\mathrm{R}}_{+ - -} G^{\mathrm{R}}_{- - -}
			G^{\mathrm{A}}_{- - -}
		\right.
	\notag \\
	& \qquad
		\left.
			+
			v_{z}
			\left( v_{x}^{+ + +} v_{y}^{+ - -} - v_{y}^{+ + +} v_{x}^{+ - -} \right)
			(G^{\mathrm{R}}_{+ + +})^{3} G^{\mathrm{R}}_{+ + -} G^{\mathrm{R}}_{+ - -}
			G^{\mathrm{A}}_{+ - -} G^{\mathrm{A}}_{- - -}
		\right.
	\notag \\
	& \qquad
		\left.
			+
			v_{z}
			\left( v_{x}^{+ + +} v_{y}^{+ + -} - v_{y}^{+ + +} v_{x}^{+ + -} \right)
			(G^{\mathrm{R}}_{+ + +})^{3} G^{\mathrm{R}}_{+ + -}
			G^{\mathrm{A}}_{+ + -} G^{\mathrm{A}}_{+ - -} G^{\mathrm{A}}_{- - -}
		\right\}
	\Biggr],
	\\
	\chi^{\mathrm{II}}
	=&
	\Im
	\Biggl[
		2 i
		\left\{ 1 + i \partial_{\varepsilon} \gamma \right\}
		\sum_{\bm{q}_{1},\bm{q}_{2},\bm{q}_{3}}
		\mathcal{S} (\bm{q}_{1},\bm{q}_{2},\bm{q}_{3})
		\mathrm{e}^{i \bm{q} \cdot \bm{r}}
	\notag \\
	& \quad
		\times
		\sum_{\bm{k}}
		\left\{ 
			v_{z}
			\left( v_{x}^{+ - -} v_{y}^{- - -} - v_{y}^{+ - -} v_{x}^{- - -} \right)
			G^{\mathrm{R}}_{+ + +} G^{\mathrm{R}}_{+ + -} (G^{\mathrm{R}}_{+ - -})^{2} (G^{\mathrm{R}}_{- - -})^{2}
			G^{\mathrm{A}}_{- - -}
		\right.
	\notag \\
	& \qquad
		\left.
			+
			v_{z}
			\left( v_{x}^{+ + -} v_{y}^{- - -} - v_{y}^{+ + -} v_{x}^{- - -} \right)
			G^{\mathrm{R}}_{+ + +} (G^{\mathrm{R}}_{+ + -})^{2} G^{\mathrm{R}}_{+ - -} (G^{\mathrm{R}}_{- - -})^{2}
			G^{\mathrm{A}}_{- - -}
		\right.
	\notag \\
	& \qquad
		\left.
			+
			v_{z}
			\left( v_{x}^{+ + -} v_{y}^{- - -} - v_{y}^{+ + -} v_{x}^{- - -} \right)
			G^{\mathrm{R}}_{+ + +} (G^{\mathrm{R}}_{+ + -})^{2} (G^{\mathrm{R}}_{+ - -})^{2} G^{\mathrm{R}}_{- - -}
			G^{\mathrm{A}}_{- - -}
		\right.
	\notag \\
	& \qquad
		\left.
			+
			v_{z}
			\left( v_{x}^{+ + -} v_{y}^{+ - -} - v_{y}^{+ + -} v_{x}^{+ - -} \right)
			G^{\mathrm{R}}_{+ + +} (G^{\mathrm{R}}_{+ + -})^{2} (G^{\mathrm{R}}_{+ - -})^{2}
			G^{\mathrm{A}}_{+ - -} G^{\mathrm{A}}_{- - -}
		\right.
	\notag \\
	& \qquad
		\left.
			+
			v_{z}
			\left( v_{x}^{+ + +} v_{y}^{- - -} - v_{y}^{+ + +} v_{x}^{- - -} \right)
			(G^{\mathrm{R}}_{+ + +})^{2} G^{\mathrm{R}}_{+ + -} G^{\mathrm{R}}_{+ - -} (G^{\mathrm{R}}_{- - -})^{2}
			G^{\mathrm{A}}_{- - -}
		\right.
	\notag \\
	& \qquad
		\left.
			+
			v_{z}
			\left( v_{x}^{+ + +} v_{y}^{- - -} - v_{y}^{+ + +} v_{x}^{- - -} \right)
			(G^{\mathrm{R}}_{+ + +})^{2} G^{\mathrm{R}}_{+ + -} (G^{\mathrm{R}}_{+ - -})^{2} G^{\mathrm{R}}_{- - -}
			G^{\mathrm{A}}_{- - -}
		\right.
	\notag \\
	& \qquad
		\left.
			+
			v_{z}
			\left( v_{x}^{+ + +} v_{y}^{+ - -} - v_{y}^{+ + +} v_{x}^{+ - -} \right)
			(G^{\mathrm{R}}_{+ + +})^{2} G^{\mathrm{R}}_{+ + -} (G^{\mathrm{R}}_{+ - -})^{2}
			G^{\mathrm{A}}_{+ - -} G^{\mathrm{A}}_{- - -}
		\right.
	\notag \\
	& \qquad
		\left.
			+
			v_{z}
			\left( v_{x}^{+ + +} v_{y}^{- - -} - v_{y}^{+ + +} v_{x}^{- - -} \right)
			(G^{\mathrm{R}}_{+ + +})^{2} (G^{\mathrm{R}}_{+ + -})^{2} G^{\mathrm{R}}_{+ - -} G^{\mathrm{R}}_{- - -}
			G^{\mathrm{A}}_{- - -}
		\right.
	\notag \\
	& \qquad
		\left.
			+
			v_{z}
			\left( v_{x}^{+ + +} v_{y}^{+ - -} - v_{y}^{+ + +} v_{x}^{+ - -} \right)
			(G^{\mathrm{R}}_{+ + +})^{2} (G^{\mathrm{R}}_{+ + -})^{2} G^{\mathrm{R}}_{+ - -}
			G^{\mathrm{A}}_{+ - -} G^{\mathrm{A}}_{- - -}
		\right.
	\notag \\
	& \qquad
		\left.
			+
			v_{z}
			\left( v_{x}^{+ + +} v_{y}^{+ + -} - v_{y}^{+ + +} v_{x}^{+ + -} \right)
			(G^{\mathrm{R}}_{+ + +})^{2} (G^{\mathrm{R}}_{+ + -})^{2} 
			G^{\mathrm{A}}_{+ + -} G^{\mathrm{A}}_{+ - -} G^{\mathrm{A}}_{- - -}
		\right.
	\notag \\
	& \qquad
		\left.
			+
			v_{z}
			\left( v_{x}^{+ - -} v_{y}^{- - -} - v_{y}^{+ - -} v_{x}^{- - -} \right)
			G^{\mathrm{R}}_{+ + +} G^{\mathrm{R}}_{+ + -} (G^{\mathrm{R}}_{+ - -})^{3} G^{\mathrm{R}}_{- - -}
			G^{\mathrm{A}}_{- - -}
		\right.
	\notag \\
	& \qquad
		\left.
			+
			v_{z}
			\left( v_{x}^{+ + -} v_{y}^{- - -} - v_{y}^{+ + -} v_{x}^{- - -} \right)
			G^{\mathrm{R}}_{+ + +} (G^{\mathrm{R}}_{+ + -})^{3} G^{\mathrm{R}}_{+ - -} G^{\mathrm{R}}_{- - -}
			G^{\mathrm{A}}_{- - -}
		\right.
	\notag \\
	& \qquad
		\left.
			+
			v_{z}
			\left( v_{x}^{+ + -} v_{y}^{+ - -} - v_{y}^{+ + -} v_{x}^{+ - -} \right)
			G^{\mathrm{R}}_{+ + +} (G^{\mathrm{R}}_{+ + -})^{3} G^{\mathrm{R}}_{+ - -}
			G^{\mathrm{A}}_{+ - -} G^{\mathrm{A}}_{- - -}
		\right.
	\notag \\
	& \qquad
		\left.
			+
			v_{z}
			\left( v_{x}^{+ + +} v_{y}^{- - -} - v_{y}^{+ + +} v_{x}^{- - -} \right)
			(G^{\mathrm{R}}_{+ + +})^{3} G^{\mathrm{R}}_{+ + -} G^{\mathrm{R}}_{+ - -} G^{\mathrm{R}}_{- - -}
			G^{\mathrm{A}}_{- - -}
		\right.
	\notag \\
	& \qquad
		\left.
			+
			v_{z}
			\left( v_{x}^{+ + +} v_{y}^{+ - -} - v_{y}^{+ + +} v_{x}^{+ - -} \right)
			(G^{\mathrm{R}}_{+ + +})^{3} G^{\mathrm{R}}_{+ + -} G^{\mathrm{R}}_{+ - -}
			G^{\mathrm{A}}_{+ - -} G^{\mathrm{A}}_{- - -}
		\right.
	\notag \\
	& \qquad
		\left.
			+
			v_{z}
			\left( v_{x}^{+ + +} v_{y}^{+ + -} - v_{y}^{+ + +} v_{x}^{+ + -} \right)
			(G^{\mathrm{R}}_{+ + +})^{3} G^{\mathrm{R}}_{+ + -}
			G^{\mathrm{A}}_{+ + -} G^{\mathrm{A}}_{+ - -} G^{\mathrm{A}}_{- - -}
		\right\}
	\Biggr],
	\\
	\chi^{\mathrm{III}}
	=&
	\Im
	\Biggl[ 
		2 i
		\left\{ 1 - i \partial_{\varepsilon} \gamma \right\}
		\sum_{\bm{q}_{1},\bm{q}_{2},\bm{q}_{3}}
		\mathcal{S} (\bm{q}_{1},\bm{q}_{2},\bm{q}_{3})
		\mathrm{e}^{i \bm{q} \cdot \bm{r}}
	\notag \\
	& \quad
		\times
		\sum_{\bm{k}}
		\left\{ 
			v_{z}
			\left( v_{x}^{+ - -} v_{y}^{- - -} - v_{y}^{+ - -} v_{x}^{- - -} \right)
			G^{\mathrm{R}}_{+ + +} G^{\mathrm{R}}_{+ + -} (G^{\mathrm{R}}_{+ - -})^{2} G^{\mathrm{R}}_{- - -}
			(G^{\mathrm{A}}_{- - -})^{2}
		\right.
	\notag \\
	& \qquad
		\left.
			+
			v_{z}
			\left( v_{x}^{+ + -} v_{y}^{- - -} - v_{y}^{+ + -} v_{x}^{- - -} \right)
			G^{\mathrm{R}}_{+ + +} (G^{\mathrm{R}}_{+ + -})^{2} G^{\mathrm{R}}_{+ - -} G^{\mathrm{R}}_{- - -}
			(G^{\mathrm{A}}_{- - -})^{2}
		\right.
	\notag \\
	& \qquad
		\left.
			+
			v_{z}
			\left( v_{x}^{+ + -} v_{y}^{+ - -} - v_{y}^{+ + -} v_{x}^{+ - -} \right)
			G^{\mathrm{R}}_{+ + +} (G^{\mathrm{R}}_{+ + -})^{2} G^{\mathrm{R}}_{+ - -}
			G^{\mathrm{A}}_{+ - -} (G^{\mathrm{A}}_{- - -})^{2}
		\right.
	\notag \\
	& \qquad
		\left.
			+
			v_{z}
			\left( v_{x}^{+ + -} v_{y}^{+ - -} - v_{y}^{+ + -} v_{x}^{+ - -} \right)
			G^{\mathrm{R}}_{+ + +} (G^{\mathrm{R}}_{+ + -})^{2} G^{\mathrm{R}}_{+ - -}
			(G^{\mathrm{A}}_{+ - -})^{2} G^{\mathrm{A}}_{- - -}
		\right.
	\notag \\
	& \qquad
		\left.
			+
			v_{z}
			\left( v_{x}^{+ + +} v_{y}^{- - -} - v_{y}^{+ + +} v_{x}^{- - -} \right)
			(G^{\mathrm{R}}_{+ + +})^{2} G^{\mathrm{R}}_{+ + -} G^{\mathrm{R}}_{+ - -} G^{\mathrm{R}}_{- - -}
			(G^{\mathrm{A}}_{- - -})^{2}
		\right.
	\notag \\
	& \qquad
		\left.
			+
			v_{z}
			\left( v_{x}^{+ + +} v_{y}^{+ - -} - v_{y}^{+ + +} v_{x}^{+ - -} \right)
			(G^{\mathrm{R}}_{+ + +})^{2} G^{\mathrm{R}}_{+ + -} G^{\mathrm{R}}_{+ - -}
			G^{\mathrm{A}}_{+ - -} (G^{\mathrm{A}}_{- - -})^{2}
		\right.
	\notag \\
	& \qquad
		\left.
			+
			v_{z}
			\left( v_{x}^{+ + +} v_{y}^{+ - -} - v_{y}^{+ + +} v_{x}^{+ - -} \right)
			(G^{\mathrm{R}}_{+ + +})^{2} G^{\mathrm{R}}_{+ + -} G^{\mathrm{R}}_{+ - -}
			(G^{\mathrm{A}}_{+ - -})^{2} G^{\mathrm{A}}_{- - -}
		\right.
	\notag \\
	& \qquad
		\left.
			+
			v_{z}
			\left( v_{x}^{+ + +} v_{y}^{+ + -} - v_{y}^{+ + +} v_{x}^{+ + -} \right)
			(G^{\mathrm{R}}_{+ + +})^{2} G^{\mathrm{R}}_{+ + -}
			G^{\mathrm{A}}_{+ + -} G^{\mathrm{A}}_{+ - -} (G^{\mathrm{A}}_{- - -})^{2}
		\right.
	\notag \\
	& \qquad
		\left.
			+
			v_{z}
			\left( v_{x}^{+ + +} v_{y}^{+ + -} - v_{y}^{+ + +} v_{x}^{+ + -} \right)
			(G^{\mathrm{R}}_{+ + +})^{2} G^{\mathrm{R}}_{+ + -}
			G^{\mathrm{A}}_{+ + -} (G^{\mathrm{A}}_{+ - -})^{2} G^{\mathrm{A}}_{- - -}
		\right.
	\notag \\
	& \qquad
		\left.
			+
			v_{z}
			\left( v_{x}^{+ + +} v_{y}^{+ + -} - v_{y}^{+ + +} v_{x}^{+ + -} \right)
			(G^{\mathrm{R}}_{+ + +})^{2} G^{\mathrm{R}}_{+ + -}
			(G^{\mathrm{A}}_{+ + -})^{2} G^{\mathrm{A}}_{+ - -} G^{\mathrm{A}}_{- - -}
		\right\}
	\Biggr],
\end{align}
and
\begin{align}
	\mathcal{S}(\bm{q}_{1},\bm{q}_{2},\bm{q}_{3})
	=&
	n^{x}(\bm{q}_{1}) n^{y}(\bm{q}_{2}) n^{z}(\bm{q}_{3})
	+
	n^{y}(\bm{q}_{1}) n^{z}(\bm{q}_{2}) n^{x}(\bm{q}_{3})
	+
	n^{z}(\bm{q}_{1}) n^{x}(\bm{q}_{2}) n^{y}(\bm{q}_{3})
	\notag \\
	&
	-
	n^{x}(\bm{q}_{1}) n^{z}(\bm{q}_{2}) n^{y}(\bm{q}_{3})
	-
	n^{z}(\bm{q}_{1}) n^{y}(\bm{q}_{2}) n^{x}(\bm{q}_{3})
	-
	n^{y}(\bm{q}_{1}) n^{x}(\bm{q}_{2}) n^{z}(\bm{q}_{3})
	\notag \\
	=&
	\bm{n}(\bm{q}_{1}) \cdot \left[ \bm{n}(\bm{q}_{2}) \times \bm{n}(\bm{q}_{3}) \right].
\end{align}
Here we define
$G^{\mathrm{R(A)}}_{\pm \pm \pm}
= G^{\mathrm{R(A)}}_{\bm{k} \pm \frac{\bm{q}_{1}}{2} \pm \frac{\bm{q}_{2}}{2} \pm \frac{\bm{q}_{3}}{2}}$
and
$v_{i}^{\pm \pm \pm} = \left( k_{i} \pm \frac{q_{1,i}}{2} \pm \frac{q_{2,i}}{2} \pm \frac{q_{3,i}}{2} \right)/m$.
We decompose the kernel $\chi^{\mathrm{I}} - \chi^{\mathrm{II}} - \chi^{\mathrm{III}}$ as
\begin{align}
	\chi^{\mathrm{I}} - \chi^{\mathrm{II}} - \chi^{\mathrm{III}}
	=
	\chi_{1}^{1} + \chi^{2}_{1} + \chi^{3}_{1}
	+
	\chi_{2}^{1} + \chi^{2}_{2} + \chi^{3}_{2},
\end{align}
with
\begin{align}
	\chi^{1}_{1}
	=&
	\Im
	\Biggl[ 
		\frac{2i}{m}
		\sum_{\bm{q}_{1},\bm{q}_{2},\bm{q}_{3}}
		\mathcal{S} (\bm{q}_{1},\bm{q}_{2},\bm{q}_{3})
		\mathrm{e}^{i \bm{q} \cdot \bm{r}}
		\sum_{\bm{k}}
		(q_{1x} v_{y} - q_{1y} v_{x}) v_{z}
	\notag \\
	& \quad
		\times
		\left\{ 
			\left( 1 + i \partial_{\varepsilon} \gamma \right)
			\left[ 
				-
				G^{\mathrm{R}}_{+ + +} G^{\mathrm{R}}_{+ + -} (G^{\mathrm{R}}_{+ - -})^{2} (G^{\mathrm{R}}_{- - -})^{2}
				G^{\mathrm{A}}_{- - -}
				-
				G^{\mathrm{R}}_{+ + +} (G^{\mathrm{R}}_{+ + -})^{2} G^{\mathrm{R}}_{+ - -} (G^{\mathrm{R}}_{- - -})^{2}
				G^{\mathrm{A}}_{- - -}
			\right.
		\right.
	\notag \\
	& \qquad \quad
		\left.
			\left.
				-
				(G^{\mathrm{R}}_{+ + +})^{2} G^{\mathrm{R}}_{+ + -} G^{\mathrm{R}}_{+ - -} (G^{\mathrm{R}}_{- - -})^{2}
				G^{\mathrm{A}}_{- - -}
			\right]
		\right.
	\notag \\
	& \qquad
		\left.
			-
			\left( 1 - i \partial_{\varepsilon} \gamma \right)
			\left[ 
				G^{\mathrm{R}}_{+ + +} G^{\mathrm{R}}_{+ + -} (G^{\mathrm{R}}_{+ - -})^{2} G^{\mathrm{R}}_{- - -}
				(G^{\mathrm{A}}_{- - -})^{2}
				+
				G^{\mathrm{R}}_{+ + +} (G^{\mathrm{R}}_{+ + -})^{2} G^{\mathrm{R}}_{+ - -} G^{\mathrm{R}}_{- - -}
				(G^{\mathrm{A}}_{- - -})^{2}
			\right.
		\right.
	\notag \\
	& \qquad \quad
		\left.
			\left.
				+
				(G^{\mathrm{R}}_{+ + +})^{2} G^{\mathrm{R}}_{+ + -} G^{\mathrm{R}}_{+ - -} G^{\mathrm{R}}_{- - -}
				(G^{\mathrm{A}}_{- - -})^{2}
			\right]
		\right\}
	\Biggr],
	\\
	\chi^{2}_{1}
	=&
	\Im
	\Biggl[ 
		\frac{2i}{m}
		\sum_{\bm{q}_{1},\bm{q}_{2},\bm{q}_{3}}
		\mathcal{S} (\bm{q}_{1},\bm{q}_{2},\bm{q}_{3})
		\mathrm{e}^{i \bm{q} \cdot \bm{r}}
		\sum_{\bm{k}}
		(q_{2x} v_{y} - q_{2y} v_{x}) v_{z}
	\notag \\
	& \quad
		\times
		\left\{ 
			\left( 1 + i \partial_{\varepsilon} \gamma \right)
			\left[ 
				-
				G^{\mathrm{R}}_{+ + +} (G^{\mathrm{R}}_{+ + -})^{2} G^{\mathrm{R}}_{+ - -} (G^{\mathrm{R}}_{- - -})^{2}
				G^{\mathrm{A}}_{- - -}
				-
				G^{\mathrm{R}}_{+ + +} (G^{\mathrm{R}}_{+ + -})^{2} (G^{\mathrm{R}}_{+ - -})^{2} G^{\mathrm{R}}_{- - -}
				G^{\mathrm{A}}_{- - -}
			\right.
		\right.
	\notag \\
	& \qquad \quad
		\left.
			\left.
				-
				G^{\mathrm{R}}_{+ + +} (G^{\mathrm{R}}_{+ + -})^{2} (G^{\mathrm{R}}_{+ - -})^{2}
				G^{\mathrm{A}}_{+ - -} G^{\mathrm{A}}_{- - -}
				-
				(G^{\mathrm{R}}_{+ + +})^{2} G^{\mathrm{R}}_{+ + -} G^{\mathrm{R}}_{+ - -} (G^{\mathrm{R}}_{- - -})^{2}
				G^{\mathrm{A}}_{- - -}
			\right.
		\right.
	\notag \\
	& \qquad \quad
		\left.
			\left.
				-
				(G^{\mathrm{R}}_{+ + +})^{2} G^{\mathrm{R}}_{+ + -} (G^{\mathrm{R}}_{+ - -})^{2} G^{\mathrm{R}}_{- - -}
				G^{\mathrm{A}}_{- - -}
				-
				(G^{\mathrm{R}}_{+ + +})^{2} G^{\mathrm{R}}_{+ + -} (G^{\mathrm{R}}_{+ - -})^{2}
				G^{\mathrm{A}}_{+ - -} G^{\mathrm{A}}_{- - -}
			\right]
		\right.
	\notag \\
	& \qquad
		\left.
			-
			\left( 1 - i \partial_{\varepsilon} \gamma \right)
			\left[ 
				G^{\mathrm{R}}_{+ + +} (G^{\mathrm{R}}_{+ + -})^{2} G^{\mathrm{R}}_{+ - -} G^{\mathrm{R}}_{- - -}
				(G^{\mathrm{A}}_{- - -})^{2}
				+
				G^{\mathrm{R}}_{+ + +} (G^{\mathrm{R}}_{+ + -})^{2} G^{\mathrm{R}}_{+ - -}
				G^{\mathrm{A}}_{+ - -} (G^{\mathrm{A}}_{- - -})^{2}
			\right.
		\right.
	\notag \\
	& \qquad \quad
		\left.
			\left.
				+
				G^{\mathrm{R}}_{+ + +} (G^{\mathrm{R}}_{+ + -})^{2} G^{\mathrm{R}}_{+ - -}
				(G^{\mathrm{A}}_{+ - -})^{2} G^{\mathrm{A}}_{- - -}
				+
				(G^{\mathrm{R}}_{+ + +})^{2} G^{\mathrm{R}}_{+ + -} G^{\mathrm{R}}_{+ - -} G^{\mathrm{R}}_{- - -}
				(G^{\mathrm{A}}_{- - -})^{2}
			\right.
		\right.
	\notag \\
	& \qquad \quad
		\left.
			\left.
				+
				(G^{\mathrm{R}}_{+ + +})^{2} G^{\mathrm{R}}_{+ + -} G^{\mathrm{R}}_{+ - -}
				G^{\mathrm{A}}_{+ - -} (G^{\mathrm{A}}_{- - -})^{2}
				+
				(G^{\mathrm{R}}_{+ + +})^{2} G^{\mathrm{R}}_{+ + -} G^{\mathrm{R}}_{+ - -}
				(G^{\mathrm{A}}_{+ - -})^{2} G^{\mathrm{A}}_{- - -}
			\right]
		\right\}
	\Biggr],
	\\
	\chi^{3}_{1}
	=&
	\Im
	\Biggl[ 
		\frac{2i}{m}
		\sum_{\bm{q}_{1},\bm{q}_{2},\bm{q}_{3}}
		\mathcal{S} (\bm{q}_{1},\bm{q}_{2},\bm{q}_{3})
		\mathrm{e}^{i \bm{q} \cdot \bm{r}}
		\sum_{\bm{k}}
		(q_{3x} v_{y} - q_{3y} v_{x}) v_{z}
	\notag \\
	& \quad
		\times
		\left\{ 
			\left( 1 + i \partial_{\varepsilon} \gamma \right)
			\left[ 
				-
				(G^{\mathrm{R}}_{+ + +})^{2} G^{\mathrm{R}}_{+ + -} G^{\mathrm{R}}_{+ - -} (G^{\mathrm{R}}_{- - -})^{2}
				G^{\mathrm{A}}_{- - -}
				-
				(G^{\mathrm{R}}_{+ + +})^{2} G^{\mathrm{R}}_{+ + -} (G^{\mathrm{R}}_{+ - -})^{2} G^{\mathrm{R}}_{- - -}
				G^{\mathrm{A}}_{- - -}
			\right.
		\right.
	\notag \\
	& \qquad \quad
		\left.
			\left.
				-
				(G^{\mathrm{R}}_{+ + +})^{2} (G^{\mathrm{R}}_{+ + -})^{2} G^{\mathrm{R}}_{+ - -} G^{\mathrm{R}}_{- - -}
				G^{\mathrm{A}}_{- - -}
				-
				(G^{\mathrm{R}}_{+ + +})^{2} G^{\mathrm{R}}_{+ + -} (G^{\mathrm{R}}_{+ - -})^{2}
				G^{\mathrm{A}}_{+ - -} G^{\mathrm{A}}_{- - -}
			\right.
		\right.
	\notag \\
	& \qquad \quad
		\left.
			\left.
				-
				(G^{\mathrm{R}}_{+ + +})^{2} (G^{\mathrm{R}}_{+ + -})^{2} G^{\mathrm{R}}_{+ - -}
				G^{\mathrm{A}}_{+ - -} G^{\mathrm{A}}_{- - -}
				-
				(G^{\mathrm{R}}_{+ + +})^{2} (G^{\mathrm{R}}_{+ + -})^{2}
				G^{\mathrm{A}}_{+ + -} G^{\mathrm{A}}_{+ - -} G^{\mathrm{A}}_{- - -}
			\right]
		\right.
	\notag \\
	& \qquad
		\left.
			-
			\left( 1 - i \partial_{\varepsilon} \gamma \right)
			\left[ 
				(G^{\mathrm{R}}_{+ + +})^{2} G^{\mathrm{R}}_{+ + -} G^{\mathrm{R}}_{+ - -} G^{\mathrm{R}}_{- - -}
				(G^{\mathrm{A}}_{- - -})^{2}
				+
				(G^{\mathrm{R}}_{+ + +})^{2} G^{\mathrm{R}}_{+ + -} G^{\mathrm{R}}_{+ - -} 
				G^{\mathrm{A}}_{+ - -} (G^{\mathrm{A}}_{- - -})^{2}
			\right.
		\right.
	\notag \\
	& \qquad \quad
		\left.
			\left.
				+
				(G^{\mathrm{R}}_{+ + +})^{2} G^{\mathrm{R}}_{+ + -} G^{\mathrm{R}}_{+ - -}
				(G^{\mathrm{A}}_{+ - -})^{2} G^{\mathrm{A}}_{- - -}
				+
				(G^{\mathrm{R}}_{+ + +})^{2} G^{\mathrm{R}}_{+ + -}
				G^{\mathrm{A}}_{+ + -} G^{\mathrm{A}}_{+ - -} (G^{\mathrm{A}}_{- - -})^{2}
			\right.
		\right.
	\notag \\
	& \qquad \quad
		\left.
			\left.
				+
				(G^{\mathrm{R}}_{+ + +})^{2} G^{\mathrm{R}}_{+ + -}
				G^{\mathrm{A}}_{+ + -} (G^{\mathrm{A}}_{+ - -})^{2} G^{\mathrm{A}}_{- - -}
				+
				(G^{\mathrm{R}}_{+ + +})^{2} G^{\mathrm{R}}_{+ + -}
				(G^{\mathrm{A}}_{+ + -})^{2} G^{\mathrm{A}}_{+ - -} G^{\mathrm{A}}_{- - -}
			\right]
		\right\}
	\Biggr],
	\\
	\chi^{1}_{2}
	=&
	\Im
	\Biggl[ 
		\frac{2i}{2m^{2}}
		\sum_{\bm{q}_{1},\bm{q}_{2},\bm{q}_{3}}
		\mathcal{S} (\bm{q}_{1},\bm{q}_{2},\bm{q}_{3})
		\mathrm{e}^{i \bm{q} \cdot \bm{r}}
		\sum_{\bm{k}}
		(- q_{2x} q_{3y} + q_{3x} q_{2y}) v_{z}
	\notag \\
	& \quad
		\times
		\left\{ 
			\left( 1 + i \partial_{\varepsilon} \gamma \right)
			\left[ 
				(G^{\mathrm{R}}_{+ + +})^{2}  (G^{\mathrm{R}}_{+ + -})^{2} G^{\mathrm{R}}_{+ - -} G^{\mathrm{R}}_{- - -}
				G^{\mathrm{A}}_{- - -}
				-
				G^{\mathrm{R}}_{+ + +} (G^{\mathrm{R}}_{+ + -})^{2} G^{\mathrm{R}}_{+ - -} (G^{\mathrm{R}}_{- - -})^{2}
				G^{\mathrm{A}}_{- - -}
			\right.
		\right.
	\notag \\
	& \qquad \quad
		\left.
			\left.
				-
				G^{\mathrm{R}}_{+ + +} (G^{\mathrm{R}}_{+ + -})^{2} (G^{\mathrm{R}}_{+ - -})^{2} G^{\mathrm{R}}_{- - -}
				G^{\mathrm{A}}_{- - -}
				+
				(G^{\mathrm{R}}_{+ + +})^{2} (G^{\mathrm{R}}_{+ + -})^{2} G^{\mathrm{R}}_{+ - -}
				G^{\mathrm{A}}_{+ - -} G^{\mathrm{A}}_{- - -}
			\right.
		\right.
	\notag \\
	& \qquad \quad
		\left.
			\left.
				-
				G^{\mathrm{R}}_{+ + +} (G^{\mathrm{R}}_{+ + -})^{2} (G^{\mathrm{R}}_{+ - -})^{2}
				G^{\mathrm{A}}_{+ - -} G^{\mathrm{A}}_{- - -}
				-
				(G^{\mathrm{R}}_{+ + +})^{2} (G^{\mathrm{R}}_{+ + -})^{2}
				G^{\mathrm{A}}_{+ + -} G^{\mathrm{A}}_{+ - -} G^{\mathrm{A}}_{- - -}
			\right]
		\right.
	\notag \\
	& \qquad
		\left.
			-
			\left( 1 - i \partial_{\varepsilon} \gamma \right)
			\left[
				G^{\mathrm{R}}_{+ + +} (G^{\mathrm{R}}_{+ + -})^{2} G^{\mathrm{R}}_{+ - -} G^{\mathrm{R}}_{- - -}
				(G^{\mathrm{A}}_{- - -})^{2}
				+
				G^{\mathrm{R}}_{+ + +} (G^{\mathrm{R}}_{+ + -})^{2} G^{\mathrm{R}}_{+ - -}
				G^{\mathrm{A}}_{+ - -} (G^{\mathrm{A}}_{- - -})^{2}
			\right.
		\right.
	\notag \\
	& \qquad \quad
		\left.
			\left.
				+
				G^{\mathrm{R}}_{+ + +} (G^{\mathrm{R}}_{+ + -})^{2} G^{\mathrm{R}}_{+ - -}
				(G^{\mathrm{A}}_{+ - -})^{2} G^{\mathrm{A}}_{- - -}
				+
				(G^{\mathrm{R}}_{+ + +})^{2} G^{\mathrm{R}}_{+ + -}
				G^{\mathrm{A}}_{+ + -} G^{\mathrm{A}}_{+ - -} (G^{\mathrm{A}}_{- - -})^{2}
			\right.
		\right.
	\notag \\
	& \qquad \quad
		\left.
			\left.
				+
				(G^{\mathrm{R}}_{+ + +})^{2} G^{\mathrm{R}}_{+ + -}
				G^{\mathrm{A}}_{+ + -} (G^{\mathrm{A}}_{+ - -})^{2} G^{\mathrm{A}}_{- - -}
				+
				(G^{\mathrm{R}}_{+ + +})^{2} G^{\mathrm{R}}_{+ + -}
				(G^{\mathrm{A}}_{+ + -})^{2} G^{\mathrm{A}}_{+ - -} G^{\mathrm{A}}_{- - -}
			\right]
		\right\}
	\Biggr],
	\\
	\chi^{2}_{2}
	=&
	\Im
	\Biggl[ 
		\frac{2i}{2m^{2}}
		\sum_{\bm{q}_{1},\bm{q}_{2},\bm{q}_{3}}
		\mathcal{S} (\bm{q}_{1},\bm{q}_{2},\bm{q}_{3})
		\mathrm{e}^{i \bm{q} \cdot \bm{r}}
		\sum_{\bm{k}}
		(- q_{1x} q_{3y} + q_{3x} q_{1y}) v_{z}
	\notag \\
	& \quad
		\times
		\left\{ 
			\left( 1 + i \partial_{\varepsilon} \gamma \right)
			\left[
				(G^{\mathrm{R}}_{+ + +})^{2} G^{\mathrm{R}}_{+ + -} (G^{\mathrm{R}}_{+ - -})^{2} G^{\mathrm{R}}_{- - -}
				G^{\mathrm{A}}_{- - -}
				-
				G^{\mathrm{R}}_{+ + +} G^{\mathrm{R}}_{+ + -} (G^{\mathrm{R}}_{+ - -})^{2} (G^{\mathrm{R}}_{- - -})^{2}
				G^{\mathrm{A}}_{- - -}
			\right.
		\right.
	\notag \\
	& \qquad \quad
		\left.
			\left.
				+
				(G^{\mathrm{R}}_{+ + +})^{2} (G^{\mathrm{R}}_{+ + -})^{2} G^{\mathrm{R}}_{+ - -} G^{\mathrm{R}}_{- - -}
				G^{\mathrm{A}}_{- - -}
				-
				G^{\mathrm{R}}_{+ + +} (G^{\mathrm{R}}_{+ + -})^{2} G^{\mathrm{R}}_{+ - -} (G^{\mathrm{R}}_{- - -})^{2}
				G^{\mathrm{A}}_{- - -}
			\right.
		\right.
	\notag \\
	& \qquad \quad
		\left.
			\left.
				-
				(G^{\mathrm{R}}_{+ + +})^{2} G^{\mathrm{R}}_{+ + -} (G^{\mathrm{R}}_{+ - -})^{2}
				G^{\mathrm{A}}_{+ - -} G^{\mathrm{A}}_{- - -}
				-
				(G^{\mathrm{R}}_{+ + +})^{2} (G^{\mathrm{R}}_{+ + -})^{2} G^{\mathrm{R}}_{+ - -}
				G^{\mathrm{A}}_{+ - -} G^{\mathrm{A}}_{- - -}
			\right.
		\right.
	\notag \\
	& \qquad \quad
		\left.
			\left.
				-
				(G^{\mathrm{R}}_{+ + +})^{2} (G^{\mathrm{R}}_{+ + -})^{2} G^{\mathrm{A}}_{+ + -}
				G^{\mathrm{A}}_{+ - -} G^{\mathrm{A}}_{- - -}
			\right]
		\right.
	\notag \\
	& \qquad
		\left.
			-
			\left( 1 - i \partial_{\varepsilon} \gamma \right)
			\left[
				G^{\mathrm{R}}_{+ + +} G^{\mathrm{R}}_{+ + -} (G^{\mathrm{R}}_{+ - -})^{2} G^{\mathrm{R}}_{- - -}
				(G^{\mathrm{A}}_{- - -})^{2}
				+
				G^{\mathrm{R}}_{+ + +} (G^{\mathrm{R}}_{+ + -})^{2} G^{\mathrm{R}}_{+ - -} G^{\mathrm{R}}_{- - -}
				(G^{\mathrm{A}}_{- - -})^{2}
			\right.
		\right.
	\notag \\
	& \qquad \quad
		\left.
			\left.
				+
				(G^{\mathrm{R}}_{+ + +})^{2} G^{\mathrm{R}}_{+ + -} G^{\mathrm{R}}_{+ - -}
				G^{\mathrm{A}}_{+ - -} (G^{\mathrm{A}}_{- - -})^{2}
				+
				(G^{\mathrm{R}}_{+ + +})^{2} G^{\mathrm{R}}_{+ + -} G^{\mathrm{R}}_{+ - -}
				(G^{\mathrm{A}}_{+ - -})^{2} G^{\mathrm{A}}_{- - -}
			\right.
		\right.
	\notag \\
	& \qquad \quad
		\left.
			\left.
				+
				(G^{\mathrm{R}}_{+ + +})^{2} G^{\mathrm{R}}_{+ + -}
				G^{\mathrm{A}}_{+ + -} G^{\mathrm{A}}_{+ - -} (G^{\mathrm{A}}_{- - -})^{2}
				+
				(G^{\mathrm{R}}_{+ + +})^{2} G^{\mathrm{R}}_{+ + -}
				G^{\mathrm{A}}_{+ + -} (G^{\mathrm{A}}_{+ - -})^{2} G^{\mathrm{A}}_{- - -}
			\right.
		\right.
	\notag \\
	& \qquad \quad
		\left.
			\left.
				+
				(G^{\mathrm{R}}_{+ + +})^{2} G^{\mathrm{R}}_{+ + -}
				(G^{\mathrm{A}}_{+ + -})^{2} G^{\mathrm{A}}_{+ - -} G^{\mathrm{A}}_{- - -}
			\right]
		\right\}
	\Biggr],
	\\
	\chi^{3}_{2}
	=&
	\Im
	\Biggl[ 
		\frac{2i}{2m^{2}}
		\sum_{\bm{q}_{1},\bm{q}_{2},\bm{q}_{3}}
		\mathcal{S} (\bm{q}_{1},\bm{q}_{2},\bm{q}_{3})
		\mathrm{e}^{i \bm{q} \cdot \bm{r}}
		\sum_{\bm{k}}
		(- q_{1x} q_{2y} + q_{2x} q_{1y}) v_{z}
	\notag \\
	& \quad
		\times
		\left\{ 
			\left( 1 + i \partial_{\varepsilon} \gamma \right)
			\left[ 
				G^{\mathrm{R}}_{+ + +} (G^{\mathrm{R}}_{+ + -})^{2} (G^{\mathrm{R}}_{+ - -})^{2} G^{\mathrm{R}}_{- - -}
				G^{\mathrm{A}}_{- - -}
				+
				(G^{\mathrm{R}}_{+ + +})^{2} G^{\mathrm{R}}_{+ + -} (G^{\mathrm{R}}_{+ - -})^{2} G^{\mathrm{R}}_{- - -}
				G^{\mathrm{A}}_{- - -}
			\right.
		\right.
	\notag \\
	& \qquad \quad
		\left.
			\left.
				-
				G^{\mathrm{R}}_{+ + +} G^{\mathrm{R}}_{+ + -} (G^{\mathrm{R}}_{+ - -})^{2} (G^{\mathrm{R}}_{- - -})^{2}
				G^{\mathrm{A}}_{- - -}
				-
				G^{\mathrm{R}}_{+ + +} (G^{\mathrm{R}}_{+ + -})^{2} (G^{\mathrm{R}}_{+ - -})^{2}
				G^{\mathrm{A}}_{+ - -} G^{\mathrm{A}}_{- - -}
			\right.
		\right.
	\notag \\
	& \qquad \quad
		\left.
			\left.
				-
				(G^{\mathrm{R}}_{+ + +})^{2} G^{\mathrm{R}}_{+ + -} (G^{\mathrm{R}}_{+ - -})^{2}
				G^{\mathrm{A}}_{+ - -} G^{\mathrm{A}}_{- - -}
			\right]
		\right.
	\notag \\
	& \qquad
		\left.
			-
			\left( 1 - i \partial_{\varepsilon} \gamma \right)
			\left[
				G^{\mathrm{R}}_{+ + +} G^{\mathrm{R}}_{+ + -} (G^{\mathrm{R}}_{+ - -})^{2} G^{\mathrm{R}}_{- - -}
				(G^{\mathrm{A}}_{- - -})^{2}
				+
				G^{\mathrm{R}}_{+ + +} (G^{\mathrm{R}}_{+ + -})^{2} G^{\mathrm{R}}_{+ - -} 
				G^{\mathrm{A}}_{+ - -} (G^{\mathrm{A}}_{- - -})^{2}
			\right.
		\right.
	\notag \\
	& \qquad \quad
		\left.
			\left.
				+
				G^{\mathrm{R}}_{+ + +} (G^{\mathrm{R}}_{+ + -})^{2} G^{\mathrm{R}}_{+ - -}
				(G^{\mathrm{A}}_{+ - -})^{2} G^{\mathrm{A}}_{- - -}
				+
				(G^{\mathrm{R}}_{+ + +})^{2} G^{\mathrm{R}}_{+ + -} G^{\mathrm{R}}_{+ - -}
				G^{\mathrm{A}}_{+ - -} (G^{\mathrm{A}}_{- - -})^{2}
			\right.
		\right.
	\notag \\
	& \qquad \quad
		\left.
			\left.
				+
				(G^{\mathrm{R}}_{+ + +})^{2} G^{\mathrm{R}}_{+ + -} G^{\mathrm{R}}_{+ - -}
				(G^{\mathrm{A}}_{+ - -})^{2} G^{\mathrm{A}}_{- - -}
			\right]
		\right\}
	\Biggr].
\end{align}
We expand $\chi^{i}_{j}$ in terms of the wavenumber of the magnetic texture $q$.
Since the contributions of the first and second order in $q$ become zero,
we focus on the third order in $q$.
First, we calculate $\chi^{i}_{1}$ ($i = 1,2,3$).
Noting the relation 
$(q_{1i} + q_{2i} + q_{3i})(q_{1j} + q_{2j} + q_{3j}) \mathcal{S}(\bm{q}_{1}, \bm{q}_{2}, \bm{q}_{3}) = 0$
(meaning $\partial_{i} \partial_{j} \left[ (\partial_{l} \bm{n}) \cdot (\bm{n} \times \bm{n}) \right] = 0$
because $\bm{n} \times \bm{n} = 0$),
we obtain
\begin{align}
	\chi^{1}_{1}
	=&
	\frac{i}{m}
	\sum_{\bm{q}_{1},\bm{q}_{2},\bm{q}_{3}}
	\left[ 
		\left\{
			(q_{1z} + q_{2z} + q_{3z}) q_{1x} (q_{1y} - q_{3y}) \mathcal{S} (\bm{q}_{1},\bm{q}_{2},\bm{q}_{3})
			+
			(y \leftrightarrow z)
		\right\}
		-
		\left\{ x \leftrightarrow y \right\}
	\right]
	\mathrm{e}^{i \bm{q} \cdot \bm{r}}
	\notag \\
	& \quad
	\times
	\Im
	\left[ 
		\sum_{\bm{k}} \frac{v^{4}}{15}
		\left\{ 
			\left( 1 + i \partial_{\varepsilon} \gamma \right)
			\left[ 
				3 (G^{\mathrm{R}})^{8} G^{\mathrm{A}}
				+
				4 (G^{\mathrm{R}})^{7} (G^{\mathrm{A}})^{2}
			\right]
			-
			\left( 1 - i \partial_{\varepsilon} \gamma \right)
			\left[ 
				(G^{\mathrm{R}})^{7} (G^{\mathrm{A}})^{2}
				-
				8 (G^{\mathrm{R}})^{6} (G^{\mathrm{A}})^{3}
			\right]
		\right\}
	\right]
	\notag \\
	&
	+ \frac{i}{m}
	\sum_{\bm{q}_{1},\bm{q}_{2},\bm{q}_{3}}
	\left[ 
		\left\{
			(q_{1z} - q_{3z}) q_{1x} (q_{1y} - q_{3y}) \mathcal{S} (\bm{q}_{1},\bm{q}_{2},\bm{q}_{3})
			+
			(y \leftrightarrow z)
		\right\}
		-
		\left\{ x \leftrightarrow y \right\}
	\right]
	\mathrm{e}^{i \bm{q} \cdot \bm{r}}
	\notag \\
	& \quad
	\times
	\Im
	\left[ 
		\sum_{\bm{k}} \frac{v^{4}}{15}
		\left\{ 
			\left( 1 + i \partial_{\varepsilon} \gamma \right)
			\left[ 
				-10 (G^{\mathrm{R}})^{8} G^{\mathrm{A}}
			\right]
			-
			\left( 1 - i \partial_{\varepsilon} \gamma \right)
			\left[ 
				10 (G^{\mathrm{R}})^{7} (G^{\mathrm{A}})^{2}
			\right]
		\right\}
	\right]
	\notag \\
	&
	+ \frac{i}{2m}
	\sum_{\bm{q}_{1},\bm{q}_{2},\bm{q}_{3}}
	\left[ 
		\left\{
			(q_{1z} + {q}_{2z} - q_{3z}) q_{1x} (q_{1y} - {q}_{2y} - q_{3y}) \mathcal{S} (\bm{q}_{1},\bm{q}_{2},\bm{q}_{3})
			+
			(y \leftrightarrow z)
		\right\}
		-
		\left\{ x \leftrightarrow y \right\}
	\right]
	\mathrm{e}^{i \bm{q} \cdot \bm{r}}
	\notag \\
	& \quad
	\times
	\Im
	\left[ 
		\sum_{\bm{k}} \frac{v^{4}}{15}
		\left\{ 
			\left( 1 + i \partial_{\varepsilon} \gamma \right)
			\left[ 
				-5 (G^{\mathrm{R}})^{8} G^{\mathrm{A}}
			\right]
			-
			\left( 1 - i \partial_{\varepsilon} \gamma \right)
			\left[ 
				5 (G^{\mathrm{R}})^{7} (G^{\mathrm{A}})^{2}
			\right]
		\right\}
	\right],
	\\
	\chi^{2}_{1}
	=&
	\frac{i}{2m}
	\sum_{\bm{q}_{1},\bm{q}_{2},\bm{q}_{3}}
	\left[ 
		\left\{
			(q_{1z} + q_{2z} + q_{3z}) q_{2x} (q_{1y} + q_{2y} - q_{3y}) \mathcal{S} (\bm{q}_{1},\bm{q}_{2},\bm{q}_{3})
			+
			(y \leftrightarrow z)
		\right\}
		-
		\left\{ x \leftrightarrow y \right\}
	\right]
	\mathrm{e}^{i \bm{q} \cdot \bm{r}}
	\notag \\
	& \quad
	\times
	\Im
	\Biggl[ 
		\sum_{\bm{k}} \frac{v^{4}}{15}
		\left\{
			\left( 1 + i \partial_{\varepsilon} \gamma \right)
			\left[
				(G^{\mathrm{R}})^{8} G^{\mathrm{A}}
				+
				2 (G^{\mathrm{R}})^{7} (G^{\mathrm{A}})^{2}
				+
				3 (G^{\mathrm{R}})^{6} (G^{\mathrm{A}})^{3}
			\right]
		\right.
	\notag \\
	& \qquad
		\left.
			-
			\left( 1 - i \partial_{\varepsilon} \gamma \right)
			\left[ 
				(G^{\mathrm{R}})^{7} (G^{\mathrm{A}})^{2}
				+
				2 (G^{\mathrm{R}})^{6} (G^{\mathrm{A}})^{3}
				-
				9 (G^{\mathrm{R}})^{5} (G^{\mathrm{A}})^{4}
			\right]
		\right\}
	\Biggr]
	\notag \\
	&
	+ \frac{i}{2m}
	\sum_{\bm{q}_{1},\bm{q}_{2},\bm{q}_{3}}
	\left[ 
		\left\{
			(q_{1z} + q_{2z} + q_{3z}) q_{2x} (q_{1y} - q_{2y} - q_{3y}) \mathcal{S} (\bm{q}_{1},\bm{q}_{2},\bm{q}_{3})
			+
			(y \leftrightarrow z)
		\right\}
		-
		\left\{ x \leftrightarrow y \right\}
	\right]
	\mathrm{e}^{i \bm{q} \cdot \bm{r}}
	\notag \\
	& \quad
	\times
	\Im
	\Biggl[ 
		\sum_{\bm{k}} \frac{v^{4}}{15}
		\left\{
			\left( 1 + i \partial_{\varepsilon} \gamma \right)
			\left[ - (G^{\mathrm{R}})^{8} G^{\mathrm{A}}
				+
				(G^{\mathrm{R}})^{6} (G^{\mathrm{A}})^{3}
				+
				2 (G^{\mathrm{R}})^{5} (G^{\mathrm{A}})^{4}
			\right]
		\right.
	\notag \\
	& \qquad
		\left.
			-
			\left( 1 - i \partial_{\varepsilon} \gamma \right)
			\left[ 
				(G^{\mathrm{R}})^{7} (G^{\mathrm{A}})^{2}
				+
				2 (G^{\mathrm{R}})^{6} (G^{\mathrm{A}})^{3}
				+
				3 (G^{\mathrm{R}})^{5} (G^{\mathrm{A}})^{4}
				-
				8 (G^{\mathrm{R}})^{4} (G^{\mathrm{A}})^{5}
			\right]
		\right\}
	\Biggr]
	\notag \\
	&
	+ \frac{i}{2m}
	\sum_{\bm{q}_{1},\bm{q}_{2},\bm{q}_{3}}
	\left[ 
		\left\{
			(q_{1z} + q_{2z} - q_{3z}) q_{2x} (q_{1y} + q_{2y} - q_{3y}) \mathcal{S} (\bm{q}_{1},\bm{q}_{2},\bm{q}_{3})
			+
			(y \leftrightarrow z)
		\right\}
		-
		\left\{ x \leftrightarrow y \right\}
	\right]
	\mathrm{e}^{i \bm{q} \cdot \bm{r}}
	\notag \\
	& \quad
	\times
	\Im
	\Biggl[ 
		\sum_{\bm{k}} \frac{v^{4}}{15}
		\left\{
			\left( 1 + i \partial_{\varepsilon} \gamma \right)
			\left[
				-16 (G^{\mathrm{R}})^{8} G^{\mathrm{A}}
				-
				8 (G^{\mathrm{R}})^{7} (G^{\mathrm{A}})^{2}
			\right]
			-
			\left( 1 - i \partial_{\varepsilon} \gamma \right)
			\left[ 
				8 (G^{\mathrm{R}})^{7} (G^{\mathrm{A}})^{2}
				+
				16 (G^{\mathrm{R}})^{6} (G^{\mathrm{A}})^{3}
			\right]
		\right\}
	\Biggr]
	\notag \\
	&
	+ \frac{i}{2m}
	\sum_{\bm{q}_{1},\bm{q}_{2},\bm{q}_{3}}
	\left[ 
		\left\{
			(q_{1z} + q_{2z} - q_{3z}) q_{2x} (q_{1y} - q_{2y} - q_{3y}) \mathcal{S} (\bm{q}_{1},\bm{q}_{2},\bm{q}_{3})
			+
			(y \leftrightarrow z)
		\right\}
		-
		\left\{ x \leftrightarrow y \right\}
	\right]
	\mathrm{e}^{i \bm{q} \cdot \bm{r}}
	\notag \\
	& \quad
	\times
	\Im
	\Biggl[ 
		\sum_{\bm{k}} \frac{v^{4}}{15}
		\left\{
			\left( 1 + i \partial_{\varepsilon} \gamma \right)
			\left[
				-9 (G^{\mathrm{R}})^{8} G^{\mathrm{A}}
				-
				6 (G^{\mathrm{R}})^{7} (G^{\mathrm{A}})^{2}
				-
				3 (G^{\mathrm{R}})^{6} (G^{\mathrm{A}})^{3}
			\right]
		\right.
	\notag \\
	& \qquad
		\left.
			-
			\left( 1 - i \partial_{\varepsilon} \gamma \right)
			\left[ 
				3 (G^{\mathrm{R}})^{7} (G^{\mathrm{A}})^{2}
				+
				6 (G^{\mathrm{R}})^{6} (G^{\mathrm{A}})^{3}
				+
				9 (G^{\mathrm{R}})^{5} (G^{\mathrm{A}})^{4}
			\right]
		\right\}
	\Biggr]
	\notag \\
	&
	+ \frac{i}{2m}
	\sum_{\bm{q}_{1},\bm{q}_{2},\bm{q}_{3}}
	\left[ 
		\left\{
			(q_{1z} - q_{2z} - q_{3z}) q_{2x} (q_{1y} - q_{2y} - q_{3y}) \mathcal{S} (\bm{q}_{1},\bm{q}_{2},\bm{q}_{3})
			+
			(y \leftrightarrow z)
		\right\}
		-
		\left\{ x \leftrightarrow y \right\}
	\right]
	\mathrm{e}^{i \bm{q} \cdot \bm{r}}
	\notag \\
	& \quad
	\times
	\Im
	\Biggl[ 
		\sum_{\bm{k}} \frac{v^{4}}{15}
		\left\{
			\left( 1 + i \partial_{\varepsilon} \gamma \right)
			\left[
				-16 (G^{\mathrm{R}})^{8} G^{\mathrm{A}}
				-
				12 (G^{\mathrm{R}})^{7} (G^{\mathrm{A}})^{2}
				-
				8 (G^{\mathrm{R}})^{6} (G^{\mathrm{A}})^{3}
				-
				4 (G^{\mathrm{R}})^{5} (G^{\mathrm{A}})^{4}
			\right]
		\right.
	\notag \\
	& \qquad
		\left.
			-
			\left( 1 - i \partial_{\varepsilon} \gamma \right)
			\left[ 
				4 (G^{\mathrm{R}})^{7} (G^{\mathrm{A}})^{2}
				+
				8 (G^{\mathrm{R}})^{6} (G^{\mathrm{A}})^{3}
				+
				12 (G^{\mathrm{R}})^{5} (G^{\mathrm{A}})^{4}
				+
				16 (G^{\mathrm{R}})^{4} (G^{\mathrm{A}})^{5}
			\right]
		\right\}
	\Biggr],
	\\
	\chi^{3}_{1}
	=&
	\frac{i}{2m}
	\sum_{\bm{q}_{1},\bm{q}_{2},\bm{q}_{3}}
	\left[ 
		\left\{
			(q_{1z} + q_{2z} + q_{3z}) q_{3x} (q_{1y} + q_{2y} - q_{3y}) \mathcal{S} (\bm{q}_{1},\bm{q}_{2},\bm{q}_{3})
			+
			(y \leftrightarrow z)
		\right\}
		-
		\left\{ x \leftrightarrow y \right\}
	\right]
	\mathrm{e}^{i \bm{q} \cdot \bm{r}}
	\notag \\
	& \quad
	\times
	\Im
	\Biggl[ 
		\sum_{\bm{k}} \frac{v^{4}}{15}
		\left\{ 
			\left( 1 + i \partial_{\varepsilon} \gamma \right)
			\left[ 
				-
				3(G^{\mathrm{R}})^{8} G^{\mathrm{A}}
				-
				2 (G^{\mathrm{R}})^{7} (G^{\mathrm{A}})^{2}
				-
				(G^{\mathrm{R}})^{6} (G^{\mathrm{A}})^{3}
				+
				2 (G^{\mathrm{R}})^{5} (G^{\mathrm{A}})^{4}
				-
				(G^{\mathrm{R}})^{4} (G^{\mathrm{A}})^{5}
			\right]
		\right.
	\notag \\
	& \qquad
		\left.
			-
			\left( 1 - i \partial_{\varepsilon} \gamma \right)
			\left[ 
				(G^{\mathrm{R}})^{7} (G^{\mathrm{A}})^{2}
				+
				2 (G^{\mathrm{R}})^{6} (G^{\mathrm{A}})^{3}
				+
				3 (G^{\mathrm{R}})^{5} (G^{\mathrm{A}})^{4}
				+
				4 (G^{\mathrm{R}})^{4} (G^{\mathrm{A}})^{5}
				-
				5 (G^{\mathrm{R}})^{3} (G^{\mathrm{A}})^{6}
			\right]
		\right\}
	\Biggr]
	\notag \\
	&
	+ \frac{i}{2m}
	\sum_{\bm{q}_{1},\bm{q}_{2},\bm{q}_{3}}
	\left[ 
		\left\{
			(q_{1z} + q_{2z} + q_{3z}) q_{3x} (q_{1y} - q_{2y} - q_{3y}) \mathcal{S} (\bm{q}_{1},\bm{q}_{2},\bm{q}_{3})
			+
			(y \leftrightarrow z)
		\right\}
		-
		\left\{ x \leftrightarrow y \right\}
	\right]
	\mathrm{e}^{i \bm{q} \cdot \bm{r}}
	\notag \\
	& \quad
	\times
	\Im
	\Biggl[ 
		\sum_{\bm{k}} \frac{v^{4}}{15}
		\left\{ 
			\left( 1 + i \partial_{\varepsilon} \gamma \right)
			\left[ 
				-
				3 (G^{\mathrm{R}})^{8} G^{\mathrm{A}}
				-
				2 (G^{\mathrm{R}})^{7} (G^{\mathrm{A}})^{2}
				-
				(G^{\mathrm{R}})^{6} (G^{\mathrm{A}})^{3}
				+
				(G^{\mathrm{R}})^{4} (G^{\mathrm{A}})^{5}
			\right]
		\right.
	\notag \\
	& \qquad
		\left.
			-
			\left( 1 - i \partial_{\varepsilon} \gamma \right)
			\left[
				(G^{\mathrm{R}})^{7} (G^{\mathrm{A}})^{2}
				+
				2 (G^{\mathrm{R}})^{6} (G^{\mathrm{A}})^{3}
				+
				3 (G^{\mathrm{R}})^{5} (G^{\mathrm{A}})^{4}
				+
				4 (G^{\mathrm{R}})^{4} (G^{\mathrm{A}})^{5}
				-
				5 (G^{\mathrm{R}})^{3} (G^{\mathrm{A}})^{6}
			\right]
		\right\}
	\Biggr]
	\notag \\
	&
	+ \frac{i}{2m}
	\sum_{\bm{q}_{1},\bm{q}_{2},\bm{q}_{3}}
	\left[ 
		\left\{
			(q_{1z} + q_{2z} - q_{3z}) q_{3x} (q_{1y} + q_{2y} - q_{3y}) \mathcal{S} (\bm{q}_{1},\bm{q}_{2},\bm{q}_{3})
			+
			(y \leftrightarrow z)
		\right\}
		-
		\left\{ x \leftrightarrow y \right\}
	\right]
	\mathrm{e}^{i \bm{q} \cdot \bm{r}}
	\notag \\
	& \quad
	\times
	\Im
	\Biggl[ 
		\sum_{\bm{k}} \frac{v^{4}}{15}
		\left\{ 
			\left( 1 + i \partial_{\varepsilon} \gamma \right)
			\left[ 
				-
				10 (G^{\mathrm{R}})^{8} G^{\mathrm{A}}
				-
				8 (G^{\mathrm{R}})^{7} (G^{\mathrm{A}})^{2}
				-
				6 (G^{\mathrm{R}})^{6} (G^{\mathrm{A}})^{3}
				-
				4 (G^{\mathrm{R}})^{5} (G^{\mathrm{A}})^{4}
				-
				2 (G^{\mathrm{R}})^{4} (G^{\mathrm{A}})^{5}
			\right]
		\right.
	\notag \\
	& \qquad
		\left.
			-
			\left( 1 - i \partial_{\varepsilon} \gamma \right)
			\left[ 
				2 (G^{\mathrm{R}})^{7} (G^{\mathrm{A}})^{2}
				+
				4 (G^{\mathrm{R}})^{6} (G^{\mathrm{A}})^{3}
				+
				6 (G^{\mathrm{R}})^{5} (G^{\mathrm{A}})^{4}
				+
				8 (G^{\mathrm{R}})^{4} (G^{\mathrm{A}})^{5}
				+
				10 (G^{\mathrm{R}})^{3} (G^{\mathrm{A}})^{6}
			\right]
		\right\}
	\Biggr]
	\notag \\
	&
	+ \frac{i}{2m}
	\sum_{\bm{q}_{1},\bm{q}_{2},\bm{q}_{3}}
	\left[ 
		\left\{
			(q_{1z} + q_{2z} - q_{3z}) q_{3x} (q_{1y} - q_{2y} - q_{3y}) \mathcal{S} (\bm{q}_{1},\bm{q}_{2},\bm{q}_{3})
			+
			(y \leftrightarrow z)
		\right\}
		-
		\left\{ x \leftrightarrow y \right\}
	\right]
	\mathrm{e}^{i \bm{q} \cdot \bm{r}}
	\notag \\
	& \quad
	\times
	\Im
	\Biggl[ 
		\sum_{\bm{k}} \frac{v^{4}}{15}
		\left\{ 
			\left( 1 + i \partial_{\varepsilon} \gamma \right)
			\left[ 
				-
				5 (G^{\mathrm{R}})^{8} G^{\mathrm{A}}
				-
				4 (G^{\mathrm{R}})^{7} (G^{\mathrm{A}})^{2}
				-
				3 (G^{\mathrm{R}})^{6} (G^{\mathrm{A}})^{3}
				-
				2 (G^{\mathrm{R}})^{5} (G^{\mathrm{A}})^{4}
				-
				(G^{\mathrm{R}})^{4} (G^{\mathrm{A}})^{5}
			\right]
		\right.
	\notag \\
	& \qquad
		\left.
			-
			\left( 1 - i \partial_{\varepsilon} \gamma \right)
			\left[ 
				(G^{\mathrm{R}})^{7} (G^{\mathrm{A}})^{2}
				+
				2 (G^{\mathrm{R}})^{6} (G^{\mathrm{A}})^{3}
				+
				3 (G^{\mathrm{R}})^{5} (G^{\mathrm{A}})^{4}
				+
				4 (G^{\mathrm{R}})^{4} (G^{\mathrm{A}})^{5}
				+
				5 (G^{\mathrm{R}})^{3} (G^{\mathrm{A}})^{6}
			\right]
		\right\}
	\Biggr]
	\notag \\
	&
	+ \frac{i}{2m}
	\sum_{\bm{q}_{1},\bm{q}_{2},\bm{q}_{3}}
	\left[ 
		\left\{
			(q_{1z} - q_{2z} - q_{3z}) q_{3x} (q_{1y} - q_{2y} - q_{3y}) \mathcal{S} (\bm{q}_{1},\bm{q}_{2},\bm{q}_{3})
			+
			(y \leftrightarrow z)
		\right\}
		-
		\left\{ x \leftrightarrow y \right\}
	\right]
	\mathrm{e}^{i \bm{q} \cdot \bm{r}}
	\notag \\
	& \quad
	\times
	\Im
	\Biggl[ 
		\sum_{\bm{k}} \frac{v^{4}}{15}
		\left\{ 
			\left( 1 + i \partial_{\varepsilon} \gamma \right)
			\left[ 
				-
				10 (G^{\mathrm{R}})^{8} G^{\mathrm{A}}
				-
				8 (G^\mathrm{R})^{7} (G^{\mathrm{A}})^{2}
				-
				6 (G^{\mathrm{R}})^{6} (G^{\mathrm{A}})^{3}
				-
				4 (G^{\mathrm{R}})^{5} (G^{\mathrm{A}})^{4}
				-
				2 (G^{\mathrm{R}})^{4} (G^{\mathrm{A}})^{5}
			\right]
		\right.
	\notag \\
	& \qquad
		\left.
			-
			\left( 1 - i \partial_{\varepsilon} \gamma \right)
			\left[ 
				2 (G^\mathrm{R})^{7} (G^{\mathrm{A}})^{2}
				+
				4 (G^{\mathrm{R}})^{6} (G^{\mathrm{A}})^{3}
				+
				6 (G^{\mathrm{R}})^{5} (G^{\mathrm{A}})^{4}
				+
				8 (G^{\mathrm{R}})^{4} (G^{\mathrm{A}})^{5}
				+
				10 (G^{\mathrm{R}})^{3} (G^{\mathrm{A}})^{6}
			\right]
		\right\}
	\Biggr].
\end{align}
To perform the integral with respect ro the wavenumber $\bm{k}$,
we use the ``formula'' (up to the second leading terms in $\tau=1/(2 \gamma)$)
\begin{align}
	&\sum_{\bm{k}}
	F(k) (G^{\mathrm{R}})^{m} (G^{\mathrm{A}})^{n}
	=
	(-1)^{m+n}
	\int \mathrm{d} \xi
	\frac{F(k_{\xi}) \nu(\xi)}{(\xi - \varepsilon - i \gamma)^{m}(\xi - \varepsilon + i \gamma)^{n}}
	\notag \\
	&\simeq
	(-1)^{m} (i)^{n+m} \pi
	\frac{(n+m-2)!}{(m-1)!(n-1)!} 
	\left[ 
		2 (F \nu)_{\varepsilon} \tau^{n+m-1}
		-i
		\frac{m-n}{n + m -2 + \delta}
		\partial_{\varepsilon}(F \nu) \tau^{n+m-2}
	\right],
\end{align}
where $F(k)$ is an arbitrary function of $k = |\bm{k}|$,
then we obtain $\chi^{i}_{1}$ ($i= 1,2,3$) as
\begin{align}
	\chi^{1}_{1}
	=&
	\frac{i \pi v^{4} \nu \tau^{8}}{30 m}
	\sum_{\bm{q}_{1},\bm{q}_{2},\bm{q}_{3}}
	\mathcal{S} (\bm{q}_{1},\bm{q}_{2},\bm{q}_{3})
	\mathrm{e}^{i \bm{q} \cdot \bm{r}}
	\notag \\
	& \times
	\biggl[
		600 
		\left[ 
			\left\{
				(q_{1z} + q_{2z} + q_{3z}) q_{1x} (q_{1y} - q_{3y}) 
				+
				(y \leftrightarrow z)
			\right\}
			-
			\left\{ x \leftrightarrow y \right\}
		\right]
	\notag \\
	& \quad
		+
		240
		\left[ 
			\left\{
				(q_{1z} - q_{3z}) q_{1x} (q_{1y} - q_{3y})
				+
				(y \leftrightarrow z)
			\right\}
			-
			\left\{ x \leftrightarrow y \right\}
		\right]
	\notag \\
	& \quad
		+
		60
		\left[ 
			\left\{
				(q_{1z} + {q}_{2z} - q_{3z}) q_{1x} (q_{1y} - {q}_{2y} - q_{3y})
				+
				(y \leftrightarrow z)
			\right\}
			-
			\left\{ x \leftrightarrow y \right\}
		\right]
	\biggr],
	\\
	\chi^{2}_{1}
	=&
	\frac{i \pi v^{4} \nu \tau^{8}}{30m}
	\sum_{\bm{q}_{1},\bm{q}_{2},\bm{q}_{3}}
	\mathcal{S} (\bm{q}_{1},\bm{q}_{2},\bm{q}_{3}) \mathrm{e}^{i \bm{q} \cdot \bm{r}}
	\notag \\
	& \times
	\biggl[
		- 600
		\left[ 
			\left\{
				(q_{1z} + q_{2z} + q_{3z}) q_{2x} (q_{1y} + q_{2y} - q_{3y}) 
				+
				(y \leftrightarrow z)
			\right\}
			-
			\left\{ x \leftrightarrow y \right\}
		\right]
		\notag \\
		& \quad
		+
		600
		\left[ 
			\left\{
				(q_{1z} + q_{2z} + q_{3z}) q_{2x} (q_{1y} - q_{2y} - q_{3y})
				+
				(y \leftrightarrow z)
			\right\}
			-
			\left\{ x \leftrightarrow y \right\}
		\right]
		\notag \\
		& \quad
		-
		480
		\left[ 
			\left\{
				(q_{1z} + q_{2z} - q_{3z}) q_{2x} (q_{1y} + q_{2y} - q_{3y})
				+
				(y \leftrightarrow z)
			\right\}
			-
			\left\{ x \leftrightarrow y \right\}
		\right]
		\notag \\
		& \quad
		+ 
		360
		\left[ 
			\left\{
				(q_{1z} + q_{2z} - q_{3z}) q_{2x} (q_{1y} - q_{2y} - q_{3y})
				+
				(y \leftrightarrow z)
			\right\}
			-
			\left\{ x \leftrightarrow y \right\}
		\right]
		\notag \\
		& \quad
		-
		480
		\left[ 
			\left\{
				(q_{1z} - q_{2z} - q_{3z}) q_{2x} (q_{1y} - q_{2y} - q_{3y})
				+
				(y \leftrightarrow z)
			\right\}
			-
			\left\{ x \leftrightarrow y \right\}
		\right]
	\biggr],
	\\
	\chi^{3}_{1}
	=&
	\frac{i \pi v^{4} \nu \tau^{8}}{30m}
	\sum_{\bm{q}_{1},\bm{q}_{2},\bm{q}_{3}}
	\mathcal{S} (\bm{q}_{1},\bm{q}_{2},\bm{q}_{3}) \mathrm{e}^{i \bm{q} \cdot \bm{r}}
	\notag \\
	& \times
	\biggl[
		- 580
		\left[ 
			\left\{
				(q_{1z} + q_{2z} + q_{3z}) q_{3x} (q_{1y} + q_{2y} - q_{3y}) 
				+
				(y \leftrightarrow z)
			\right\}
			-
			\left\{ x \leftrightarrow y \right\}
		\right]
		\notag \\
		& \quad
		- 300 
		\left[ 
			\left\{
				(q_{1z} + q_{2z} + q_{3z}) q_{3x} (q_{1y} - q_{2y} - q_{3y})
				+
				(y \leftrightarrow z)
			\right\}
			-
			\left\{ x \leftrightarrow y \right\}
		\right]
		\notag \\
		& \quad
		+ 120
		\left[ 
			\left\{
				(q_{1z} + q_{2z} - q_{3z}) q_{3x} (q_{1y} + q_{2y} - q_{3y})
				+
				(y \leftrightarrow z)
			\right\}
			-
			\left\{ x \leftrightarrow y \right\}
		\right]
		\notag \\
		& \quad
		+ 60
		\left[ 
			\left\{
				(q_{1z} + q_{2z} - q_{3z}) q_{3x} (q_{1y} - q_{2y} - q_{3y})
				+
				(y \leftrightarrow z)
			\right\}
			-
			\left\{ x \leftrightarrow y \right\}
		\right]
		\notag \\
		& \quad
		+ 120
		\left[ 
			\left\{
				(q_{1z} - q_{2z} - q_{3z}) q_{3x} (q_{1y} - q_{2y} - q_{3y})
				+
				(y \leftrightarrow z)
			\right\}
			-
			\left\{ x \leftrightarrow y \right\}
		\right]
	\biggr].
\end{align}
After some algebla, we obtain 
\begin{align}
	\chi^{1}_{1} + \chi^{2}_{1} + \chi^{3}_{1}
	=
	\frac{20 \pi v^{4} \nu \tau^{8}}{m}
	\bm{n} \cdot \partial_{z} \left( \partial_{x} \bm{n} \times \partial_{y} \bm{n} \right).
\end{align}
On the other hand, we can easily see that $\chi^{i}_{2}$ ($i = 1,2,3$)
have the terms of order $\mathcal{O}(\tau^{7})$,
and we can neglect the $\chi^{i}_{2}$s since we assume $\varepsilon_{\mathrm{F}} \tau \gg 1$.
Therefore, we obtain the NCTE Hall current as
\begin{align}
	j^{\mathrm{NCTE}}_{z}
	=
	\int \mathrm{d} \varepsilon \left( - \frac{\partial f}{\partial \varepsilon} \right)
	(\varepsilon - \mu)
	D(\varepsilon)
    \int \frac{\mathrm{d}\bm{r}}{V}
    \bm{n} \cdot \partial_{z} \left( \partial_{x} \bm{n} \times \partial_{y} \bm{n} \right),
\end{align}
where $D(\varepsilon) = 20 e^{2} \varepsilon^{2} \nu M^{3} \tau^{8}/m^{3}$ and $V$ is the volume of the system.
We rewrite the integral
\begin{align}
    \int \frac{\mathrm{d}\bm{r}}{V} 
    \bm{n} \cdot \partial_{z} (\partial_{x} \bm{n} \times \partial_{y} \bm{n})
    =
    \int \frac{\mathrm{d}\bm{r}}{V} 
    \partial_z \left[ \bm{n} \cdot (\partial_{x} \bm{n} \times \partial_{y} \bm{n}) \right]
    -
    \int \frac{\mathrm{d}\bm{r}}{V} (\partial_{z} \bm{n}) \cdot (\partial_{x} \bm{n} \times \partial_{y} \bm{n}),
\end{align}
and introducing $b(\bm{r})$ as $\partial_{x} \bm{n} \times \partial_{y} \bm{n} = b(\bm{r}) \bm{n}$ (same definition in main text),
then the second term becomes $(\partial_{z} \bm{n}) \cdot (\partial_{x} \bm{n} \times \partial_{y} \bm{n}) = b(\bm{r}) [(\partial_{z} \bm{n}) \cdot \bm{n}] = 0$.
Therefore, we finally obtain the following result.
\begin{align}
	j^{\mathrm{NCTE}}_{z}
	=
	\int \mathrm{d} \varepsilon \left( - \frac{\partial f}{\partial \varepsilon} \right)
	(\varepsilon - \mu)
	D(\varepsilon)
    \int \frac{\mathrm{d}\bm{r}}{V}
    \partial_{z}
    \left[ 
        \bm{n} \cdot \left( \partial_{x} \bm{n} \times \partial_{y} \bm{n} \right)
    \right].
\end{align}


\twocolumngrid

\bibliography{NCTEmag}
\clearpage
\end{document}